\documentclass[journal]{IEEEtran}
\usepackage{amsmath,amssymb,graphicx,subfigure}

\usepackage{amsmath,amssymb}
\usepackage{cite}
\newcommand{\citep}{\cite}
\newcommand{\citet}{\cite}

\begin{document}
%\journal{Neuroimage}

%\begin{frontmatter}



%  \title{Summarizing the Percent Overlap of Activation in Multiple
%  fMRI Studies}    
\title{A Re-defined and Generalized Percent-Overlap-of-Activation
  Measure for Studies of fMRI Reproducibility and its Use in Identifying
  Outlier Activation Maps}


\author{Ranjan Maitra\thanks{R. Maitra is with the Department of
    Statistics, Iowa State University, Ames IA   50011-1210, USA.}}
\maketitle

\begin{abstract}
Functional Magnetic Resonance Imaging~(fMRI) is a popular non-invasive
modality to investigate activation in the human brain. The end result
of most fMRI experiments is an activation map corresponding to the
given paradigm. These maps can vary greatly from one study to the
next, so quantifying the reliability of identified activation over
several fMRI studies is important. The percent overlap of
activation~\citep{romboutsetal98,machielsenetal00} is a global
reliability measure between activation maps drawn from 
any two fMRI studies. A slightly modified but more intuitive measure
is provided by the ~\citet{jaccard1901} coefficient of 
similarity, whose use we study in this paper. A generalization of these 
measures is also proposed to comprehensively summarize the
reliability of multiple fMRI studies. Finally, a testing mechanism to
flag  potentially anomalous studies is developed.  The methodology is
illustrated on   
studies involving left- and right-hand  motor task paradigms performed
by a right-hand dominant male subject several times over a period of
two months, with excellent results.
% Although presented here in the context of multi-session
            %single subject data, the methodology is general enough to
            %apply to multi-subject data acquired under a common
            %paradigm. 
\end{abstract}


%\end{frontmatter}


%\section{}
%\label{}

\section{Introduction}
\label{introduction}
The past two decades have seen the widespread adoption of functional
Magnetic Resonance Imaging~(fMRI) as a noninvasive  tool for 
understanding human cognitive and motor functions.
% There is a wide range of radiologic tools available to researchers
% interested in imaging areas of cerebral activation. Two prominent
% imaging modalities are positron emission
% tomography~(PET)~\cite{mazziotta86} and
% electroencephalography~(EEG)~\cite{epstein83}. However, both are
% invasive and severely limit the degree of repeatability in studies
% with a given paradigm. On  the other hand, functional Magnetic
% Resonance Imaging~(fMRI)~\cite{ogawa92} is both   non-invasive and
% has higher spatial resolution, which perhaps substantially explains
% its increasing popularity in brain-mapping over the last fifteen
% years or so~\cite{kwong92}. Additionally, while it has lower
% temporal resolution relative to EEG, recent advances in  acquisition
% techniques in single-trial studies have shown promise in closing
% this gap~\cite{rosen98}. 
The primary objective of fMRI is the identification of cerebral
regions that are activated by a given stimulus or while performing
some task. Accurate identification of such voxels is however
challenged by factors such as scanner variability, potential inherent
unreliability of the MR signal, between-subject variability, subject
motion -- whether voluntary, involuntary or
stimulus-correlated~\citep{biswaletal96,genoveseetal97,hajnaletal94}
-- or the several-seconds 
delay in the onset of the blood-oxygen-level-dependent~(BOLD) response
as a result of the passage of the neural stimulus through the 
hemodynamic filter~\citep{maitraetal02}. Since most 
signal differences between activated and control or resting states are
small, typically no more than 5\%~\citep{chenandsmall07}, there is
strong possibility of identifying %apparent signal changes can be
                                %induced by motion at the  sub-pixel
                                %level resulting in the detection of 
false
positives. This lack of %reproducibility and 
reliability is disconcerting~\citep%see, for instance,
{buchsbaumetal05,derrfussetal05,ridderinkhofetal04,uttal01}), so
fMRI data are subject to pre-processing such as the removal of 
flow artifacts by digital monitoring and
filtering~\citep{biswaletal96} or  image registration to align time 
course image sequences to sub-pixel accuracy~\citep{woodetal98}. 
The quality of acquired fMRI data is only partially improved
by such pre-processing: identified activation regions still vary from
one study to the other. Quantifying the reliability of fMRI studies is
therefore needed for drawing accurate conclusions
~\citep{mcgonigleetal00,nolletal97,weietal04} and is usually done  
by calibrating repeatability of activation results across multiple
studies.  
%The objective is to describe the extent to which activation is
%consistently identified in multiple fMRI images.

There are two main approaches to quantitating reliability of activation. 
% But before proceeding further, I specify that I use the term
% ``replication'' to denote the repetition of the task or experimental
% condition to study variability. These replications are necessarily
% independent and, in the context of single-subject studies, occur on
% different scanning sessions, reasonably separated in time.  
The first involves the analysis of fMRI data that are acquired in
one or more groups of subjects performing tasks at different
time-points~(called {\em experimental replications}) or under multiple
stimulus or task-performance levels~({\em  experimental
conditions}). %The objective of defining reliability measures in these 
        %settings is to determine where the effect of the stimulus is
        %larger than subject-to-subject variation. 
The intra-class
correlation~(ICC)~\citep{shroutandfleiss79,koch82,mcgrawandwong96}
provides a measure of correlation or conformity between regions
identified as activated in multiple subjects under two or more
experimental replications and/or conditions~% quantifies reliability
                                % of activation in this framework by
                                % using % voxels identified as
                                % activated in each subject after
                                % thresholding % separately for each
                                % combination of experimental
                                % condition and % subject 
\citep{aronetal04,fernandezetal03,friedmanetal08,manoachetal01,miezinetal00,raemekersetal07,sprechtetal03}. %Note
                                %that we do not need independence
                                %among the experimental replications
                                %to calculate the ICC (thus, date from
                                %runs within a session can be
                                %used). However, formal statistical
                                %inference methods as outlined
                                %in~\citet{raemekersetal07} assume
                                %independence between  replications. 
%(Aron et al., 2006; Fern\'{a}ndez et al., 2003; Friedman et al.,
%2008; Manoach et al., 2001; Miezin et al. 2000; Raemekers et al.,
%2007, Sprecht et al., 2003). 
\citet{raemekersetal07} also recently proposed a
within- and between-measurements ICC for multi-subject studies with
multiple experimental conditions. By design however, the ICC can not
be used to determine reliability of activation in single-subject
studies with several replications. 
% \footnote{ I use the term ``replication'' to denote the repetition
% of the task or experimental condition to study variability. These
% replications are necessarily independent and, in the context of
% single-subject studies, occur on different scanning sessions,
% reasonably separated in time.} or in multi-subject studies involving
% only one experimental condition and  no replications. 

The second scenario, which is the subject of this paper, is when
replicated fMRI data are acquired on the same subject under the same
experimental condition or under multiple subjects under the same
experimental paradigm. In these scenarios,~\citet{romboutsetal98} and
\citet{machielsenetal00} have
proposed a global reliability measure for any pair of fMRI
studies: 
% Before proceeding with the definition of the percent
% overlap of activation, I note that I define it here in
% the test-retest case, but it is also applicable to the
% case where registered fMRI data are acquired on
% multiple subjects under the same experimental paradigm.
For any two replications (say, $j$ and 
$l$), the percent overlap of activation is defined as  
$\omega_{j,l} = 2V_{j,l}/(V_j + V_l)$, where $V_{j,l}$ is the number of
three-dimensional image voxels identified as activated in both the
$j$th and the $l$th replications, and
$V_j$ and $V_l$ represent the 
number of voxels identified as activated in the
$j$th and the $l$th experiments, respectively. Thus, 
$\omega_{j,l}$ is a ratio of the number of voxels identified as
activated in both replications to the average number of voxels
identified as activated in each replication. Note that $0 \leq
\omega_{j,l}\leq 1$, spanning the cases 
measuring zero to perfect overlap in identified activation at the two
ends of the scale.
 \citet{raemekersetal07} have proposed significance tests on the
percent overlap using Fisher's $z$-transformation $z'= \tanh^{-1}
\omega_{j,l}$, however, the basis for either the transformation (of a
proportion rather than correlation coefficient) or the significance 
test (which tests for the null hypothesis that 
$\omega_{j,l}=0$) in this framework is unclear. 

 A reviewer for this paper has pointed out that $\omega_{j,l}$ is
 really identical to the~\citet{dice45} or the~\citet{sorensen48}
 similarity coefficient. As such, it has been well-studied in many
 applications and found to possess the  undesirable property known  as
 ``aliasing''~\citep{tulloss97}, {\em 
   i.e.} different input values can result in values that are very
similar to one other. \citet{ruddelletal07} also found the
~\citet{jaccard1901} similarity coefficient to be the best among a
range of similarity indices in the context of comprehensively
measuring social stability. Additionally, its complement from unity
is a true distance metric~\citep{levandowskyandwinter71}. 
  % which is that it may take values for a variety of  input values.
 % This index has been all the
 % \citet{tulloss97}. \citep{ruddelletal07,colwellandcoddington94,allisonetal95} 
This paper, therefore, introduces and studies  %There are three
                                %issues relating to 
                                %the percent overlap of 
                                %activation that I address in this
                                %paper. The first is with 
its use in quantifying fMRI reproducibility~in Section~\ref{moa}. This
measure, although a slight modification to the~\citet{romboutsetal98}
and~\citet{machielsenetal00} definition of $\omega_{j,l}$, is seen to
be both intuitive and physically interpretable. %It also summarizes these 
                                %overlap measures for $M (> 2)$
                                %studies. 
At the same time, like $\omega_{j,l}$, it is also a 
                                %is a 
pairwise reliability measure so that we get $M\choose2$ overlap measures
$\omega_{j,l}, 1\leq l\leq j \leq M$ from $M$ fMRI studies. There is
no obvious way to combine these into a single, easily understood
measure of activation reliability. In Section~\ref{summary}, 
I develop a way to describe these $M\choose2$ overlap measures, using
a spectral decomposition of the matrix of these overlap measures to
arrive at a interpretable summary. %The 
                                %method is  also interpretable and
                                %reduces to the pairwise measure when
                                %$M=2$.
This is followed by a novel use of the summarized overlap measure in
flagging outliers among the $M$ studies, for which a testing strategy
is proposed in Section~\ref{outliers}. Accounting for such outliers in
inference can provide more accurate determination of activated 
regions over several studies. As opposed to the exploratory approaches
to outlier detection proposed by
~\citet{kherifetal03},~\citet{luoandnichols03} or
\citet{seghieretal07}, my testing strategy is more formal and
supplements the approaches of~\citet{mcnameeandlazar04}
or~\citet{woolrich08}. Section~\ref{experiments} demonstrates
the methodology of Section~\ref{statistics} on two sets
of experiments involving motor paradigms that were 
replicated on the same subject twelve times  over 
the course of two months.  The paper concludes with some discussion.   
\section{Statistical Methodology}
\label{statistics}
\subsection{The Jaccard Similarity Coefficient as a Modified Percent Overlap of Activation}
\label{moa}
 Define the modified  percent overlap of activation between any two
fMRI studies ($j$ and $l$) as
\begin{equation}
{}_m\omega_{j,l} = \frac{ V_{j,l}}{V_j + V_l - V_{j,l}}
\label{eqn.moa}
\end{equation}
where $V_j$, $V_l$ and $V_{j,l}$ are as before. The measure
${}_m\omega$  
has a set-theoretic interpretation: specifically, 
it is the proportion of voxels identified as activated
in both the $l$th and $j$th replications among the ones that have been
identified as activated in either. As such, it is analogous to
the~\citet{jaccard1901} similarity coefficient. Further, it can also
be viewed as the 
conditional probability that a voxel is identified as activated in
both the $l$th and the $j$th replications, given that it is identified as
activated in at least one of the two replications. There is thus a
more natural justification for defining $_m\omega_{jl}$ than there is
for $\omega_{jl}$.

\begin{figure}
\includegraphics[width=\linewidth]{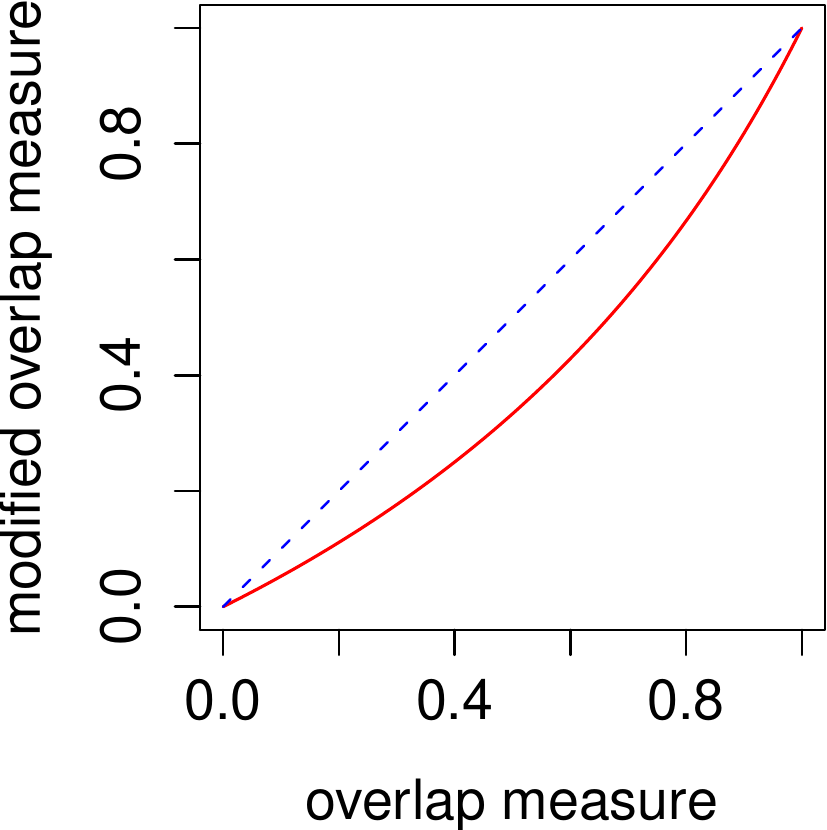} 
\caption{Plot of the modified percent overlap measure of
  activation~(Jaccard similarity coefficient)
  ${}_m\omega$ (solid red line) against the percent overlap measure of
  activation~(Dice coefficient) $\omega$. The half-broken blue line is
  the straight line 
  passing through the origin and (1, 1).}
\label{overlap-relation}
\end{figure}
Both $\omega_{j,l}$ and ${}_m\omega_{j,l}$ apply to single-subject
test-retest studies as well as to cases where registered fMRI data are
acquired on multiple subjects under the same experimental
paradigm. Further $0\leq \omega_{j,l}, {}_m\omega_{j,l}\leq 1$ 
 with $\omega_{j,l} = {}_m\omega_{j,l} =0 $
when $V_{j,l}=0$ ({\em i.e.}, no voxels activated in both
replications) and $\omega_{j,l} = {}_m\omega_{j,l} = 
1$ when $V_{j,l} = V_j = V_l$ ({\em i.e.}, the same voxels are
identified as activated in the $j$th and $l$th replications). However,
$\omega_{j,l}$ and ${}_m\omega_{j,l}$ share a non-linear relationship
between 0 and 1. To see this, note that dividing both the numerator
and denominator in~(\ref{eqn.moa}) by $2V_{j,l}$ for $V_{j,l}\neq
0$ yields
\begin{equation*}
{}_m\omega_{j,l} %& = \frac{ V_{j,l}}{V_j + V_l - V_{j,l}}\\
%& = \frac{\frac{ V_{j,l}}{2V_{j,l}}}{\frac{V_j + V_l -
%V_{j,l}}{2V_{j,l}}} \\
%& = \frac{\frac12}{\frac{V_j + V_l}{2V_{j,l}} - \frac12}\\
%& = \frac{\frac12}{\frac1{\omega_{j,l}} - \frac12}\\ & 
= \frac{\omega_{j,l}}{2-\omega_{j,l}}.
\end{equation*}
Since ${}_m\omega_{j,l} = 
\omega_{j,l} = 0$ when $V_{j,l}=0$, this 
relationship holds always. Also, $\omega_{j,l}\leq 1$, so
that $0\leq {}_m\omega_{j,l} \leq \omega_{j,l}\leq 1$, with equality
only when both are zero or unity. 
 Figure~\ref{overlap-relation} shows the relationship between
${}_m\omega_{j,l} $ and $ \omega_{j,l}$. Both
${}_m\omega_{j,l} $ and $ \omega_{j,l}$ climb from a minimum value of
zero to a maximum value of unity, but they do so at different
rates.  I next contend through two  illustrative example scenarios
that ${}_m\omega_{j,l}$ provides a more natural quantification of
overlap. All my examples  are on putative registered activation maps
of dimensions $128\times128\times 22$ voxels: thus they contain
360,448 voxels. Further, my examples are chosen to have between only
1--3.7\% of active voxels in any replication, to mimic the often
low rate of active voxels in an imaging study.   
%\paragraph{ Illustrative Example 1} In this example, there are 6,248
%(1.73\%) activated voxels in Replication A, 5,767 (1.6 \%) activated
%voxels in Replication B, and 5,407 (1.5\%) activated voxels are
%common to both  replications. Thus, there are a total of
%$6,248+5,767-5,407=6,608$ (1.83\%) voxels that are active in at least
%one of the two replications. Now, if our basis is those 6,608 voxels,
%we might be interested in finding our the proportion of these voxels
%in the overlapping area {\em i.e.}, the 5,407 common active
%voxels. Now, $5,407/6,608 = 0.818$ (our ${}_m\omega = 0.818$) makes
%intuitive sense. This contrasts with the value of $\omega = 0.90$.  

\paragraph{ Illustrative Example 1}
Replication A has 3,604 (1\%) activated voxels while Replication B
has 10,813 (3\%) activated voxels, %Thus, these two replications thus
                                %represent cases with modest to
                                %moderate activation. 
with 1,081~(0.3\%) voxels commonly identified as activated in both
replications, so there are $3,604 + 10,813 - 1,081 =  
13,336$ (3.7\%) voxels activated in at least
one of the two replications. If our basis were these 13,336 voxels,
a natural measure of coincidence is the % we might be interested in
                                % finding out  
proportion of these 
voxels in the overlapping area {\em i.e.}, the 1,081 %common active
voxels. Now $1,081/13,336 = 0.081$ which is exactly ${}_m\omega$. This
measure is therefore more intuitive than $\omega=0.150$.    

\paragraph{ Illustrative Example 2} 
This example has the same number of activated voxels for both
replications as before, but there are 3,243 (0.9\%)
voxels commonly identified as activated in both replications. 
Here ${}_m\omega = 0.29$ while $\omega = 0.45$.  The value of $\omega$
is three times that of the previous example which may, at first
glance, seem appropriate -- after all, there are three 
times more common activated voxels -- but the number of commonly
active voxels, is over a basis of far fewer voxels (11,174)
% identified as activated in at least one replication 
than previously (13,336). Thus, there is far more reliability in
activation than three times the previous value, as suggested by
$\omega$: ${}_m\omega$ which at 0.29 is 3.625 times the corresponding
value in the previous example, provides a better quantification of the 
relative sense of this reliability. 

In this section, I have introduced the Jaccard similarity coefficient
as a modified 
measure of the percent 
overlap of activation. I now introduce a generalized measure to
summarize several percent-overlap- of-activation measures.  

\subsection{Summarizing Several Pairwise Overlap Measures}
\label{summary}
Suppose we have $M$ activation maps, each obtained from a fMRI study
under the same experimental paradigm. Define
${}_m\omega_{j,j} = \omega_{j,j}=1$ for $j=1,2,\ldots,M$. For each
pair $(j, l); 1\leq j, l \leq M$ of studies, let $\omega_{j,l}$ be the
percent overlap of activation and ${}_m\omega_{j,l}$ be its
corresponding modified  version.  Further, let $\Omega = \left(\left(
    \omega_{j,l}\right ) \right)_{j=1,2,\ldots, M; l = 1,2,\ldots M} $
and ${}_m\Omega = \left(\left( {}_m\omega_{j,l}\right )
\right)_{j=1,2,\ldots, M; l = 1,2,\ldots M} $ be the matrices of the
corresponding $\omega_{j,l}$s and ${}_m\omega_{j,l}$s. These
${}_m\omega_{j,l}$s and $\omega_{j,l}$s are all
pairwise overlap measures which need to be summarized. Before
proceeding further, I note that a generalized overlap measure between
all studies could be defined in terms of the 
proportion of the voxels identified as activated in all replications
out of those identified as activated in at least one replication; but,
given the low rate of active voxels and high variability in activation
in many fMRI studies,  
this measure would in many cases be too small to be of much practical
value. ~\citet[page 122]{colwellandcoddington94} have passed on a
suggestion made by E. C. Pielou (in personal communication to them) in
deriving a 
multiple Jaccard index 
from the pairwise indices: for large $M$, this suggestion takes a
maximum value of $MS_{M}/4$, where $S_{M}$ (in our case) is the total
number of voxels activated in at least one study. This generalization
does not provide us with a proper sense for what constitutes a high or
a low value even when $M$ is large, since the upper value depends on
$S_M$ which can change from one set of studies to the next. For
more modest-sized $M$, the maximum possible attained value is not
known so that quantification is an even bigger issue. I therefore
propose to derive a measure summarizing the matrix 
of pairwise overlaps. To fix 
ideas, I develop methodology here using ${}_m\Omega$ but emphasize that
derivations are analogous for $\Omega$.

Before deriving a summarized measure over all $M$ studies, I note that   
there is highest reliability between them when %there is perfect
                                %overlap between all  pairwise maps:
                                %thus 
${}_m\omega_{j,l}=1$ for all $j,l\in \{1, 2, \ldots, M\}$% and
                                %${}_m\Omega = J_M$
, the $M\times M$-matrix of ones. On the other hand, 
the worst case is when there is zero pairwise overlap between any
two fMRI maps, then ${}_m\Omega = I_M$, the $M\times M$ identity
matrix. Any
summarized measure should assess these best- and worst-case scenarios
at the two ends of the scale. I now proceed with my derivations. 

Let ${}_m\lambda_{(1)} \geq {}_m\lambda_{(2)} \geq \ldots \geq
{}_m\lambda_{(M)}$ be the 
eigenvalues of ${}_m\Omega$; these values are all real since 
 ${}_m\Omega$ is a symmetric matrix. Further, the trace of
${}_m\Omega$ is $M$, hence ${}_m\lambda_{(1)} > 0$. I define the
summarized measure of the modified percent overlap of activation, {\em
  i.e.}, the {\em summarized multiple Jaccard similarity coefficient}
as  
\begin{equation}
{}_m^s\omega =
\frac1{M-1}\left(M\frac{{}_m\lambda_{(1)}}{\sum_{i=1}^M{}_m\lambda_{(i)}} -
  1\right) \equiv \frac{{}_m\lambda_{(1)} - 1}{M-1}, 
\label{summarized.overlap}
\end{equation}
where the last equality follows from the fact that the trace of ${}_m\Omega$
is also equal to $\sum_{i=1}^M {}_m\lambda_{(i)}$. 
The motivation for this derivation comes from principal components
analysis~(PCA) in statistics. In PCA, the ratio of the largest
eigenvalue to that of the sum of all the eigenvalues measures the
proportion of variation explained by 
the first principal component~(PC). The first PC is that %orthogonal
projection of the data which captures the maximum amount of
variability in the $M$ coordinates. PCA can be performed by obtaining
a spectral decomposition of either the correlation or the covariance
matrix, with differing results: since ${}_m\Omega$ has the flavor of a
correlation matrix with unity on the diagonals and off-diagonal
(nonnegative) elements of less than unity, I motivate ${}_m^s\omega$ 
using the analogue to the correlation matrix.  When the
correlation matrix is identity,   the coordinates are all independent
and the first PC captures
only $1/M$ fraction of variability in the data. On the other
hand, when the correlation matrix is equal to $J_M$, then all the
information is carried in one coordinate: thus the first PC explains
100\% of the variation in the data. %Like a correlation matrix,
                                %${}_m\Omega$ is also positive
                                %semi-definite with diagonal elements
                                %given by unity. 
When ${}_m\Omega = I_M$, all eigenvalues are the same so that the
ratio of the largest eigenvalue ${}_m\lambda_{(1)}$ to the sum of the
eigenvalues is equal to $1/M$. This is the worst-case scenario, so
we shift and scale the value such that the summarized measure is
zero. Alternatively, when ${}_m\Omega = J_M$, ${}_m\lambda_{(1)} =
M$ and all other eigenvalues are zero. %Thus the ratio of the largest
                                %eigenvalue ${}_m\lambda_{(1)}$ to the
                                %sum of all eigenvalues is $1$. 
This is the best-case scenario, with perfect overlap
between all replications, so the summary should take its highest
possible value of 1: ${{}_m\lambda_{(1)}}/{\sum_{i=1}^M
  {}_m\lambda_{(i)}}$ is shifted and scaled to equal 1. Solving the
two simultaneous equations for the best and worst cases yields the 
proposed summarized measure ${}_m^s\omega$ in~(\ref{summarized.overlap}). 

Note that ${}_m\Omega$ is a nonnegative symmetric matrix, {\em i.e.}
all entries are nonnegative. By the Perron-Frobenius theorem for
nonnegative matrices~(see Theorems 1.4.4 and 1.7.3 in \citep{bapatandraghavan97}), 
${}_m\lambda_{(1)}$ (which is also the {\em spectral radius} of
${}_m\Omega$) is bounded above and below by the minimum and the
maximum of the row sums, respectively. All row sums of ${}_m\Omega$
are not less than 1 (since the diagonal elements are 1) and not
greater than $M$ (since each of the $M$ elements in a row is between 0
and 1). Thus,  $0 \leq{ }^s_m\omega \leq 1$. Further, when there 
are only two replications ({\em i.e.} $M=2$), ${}_m\Omega$
is a $2\times 2$-matrix with diagonal 
elements given by unity and off-diagonal elements given by
${}_m\omega_{1,2}\equiv {}_m\omega$. Trivial algebra shows that 
${}_m\lambda_{(1)} = 1+{}_m\omega$, so ${}_m^s\omega$
in~(\ref{summarized.overlap}) reduces to ${}_m\omega$  
for $M=2$. Thus, the proposed ${}_m^s\omega$ is
consistent with the pairwise overlap measure for $M=2$.

%I conclude this section by stating that an analogous definition  as
%in~(\ref{summarized.overlap})  applies for the case of the
%(unmodified) percent overlap of activation. Specifically, letting
%$\lambda_{(1)} \geq \lambda_{(2)} \geq \ldots \geq \lambda_{(M)}$ be
%the eigenvalues of $\Omega$, a summarized measure of the  percent
%overlap of activation as   \begin{equation*} ^s\omega =
%\frac1{M-1}\left(M\frac{\lambda_{(1)}}{\sum_{i=1}^M\lambda_{(i)}} -
%1\right) \equiv \frac{\lambda_{(1)} - 1}{M-1}. \end{equation*} The
%motivation for and properties of $^s\omega$ are similar to those for
%$_m^s\omega$ and are omitted. 

\subsection{Identifying Outlying and Anomalous Activation Maps}
\label{outliers}
This section develops a testing tool using ${}_m^s\omega$  
in order to identify studies that are anomalous or outliers. As before,
$^s_m\omega$ is calculated from the $M$ studies. Further, for each
$j=1,2,\ldots,M$, let 
$^s_m\omega_{-j}$ be the summarized overlap measure obtained from the
$M-1$ studies with the $j$th study deleted. If the $j$th study is not
very similar to the other studies, {\em i.e.} $_m\omega_{j,l}$
is low for $l\neq j$, then including it %the $j$th study 
should result in
a much lower $^s_m\omega$ than $^s_m\omega_{-j}$. I propose 
a testing scheme to flag such studies.  

To do so, I advocate using as my measure 
\begin{equation*}
\zeta_{-j} = \frac 2\pi \arcsin\sqrt{{}_m^s\omega_{-j}} - \frac 2\pi \arcsin\sqrt{{}_m^s\omega}.
\end{equation*}
My motivation for applying the arcsine transformation on the
${}_m^s\omega_{-j}$ arises from the variance-stabilizing
transformation $\psi(p) = \frac 2\pi \arcsin\sqrt{p}$ often used to
approximate the distribution of the proportion of success in binomial
trials using a constant-variance normal distribution. Because $0 \leq
^s_m\!\!\omega \leq 1$, we have a similar framework as a
proportion. However, the distributional assumption governing the form
of $\zeta_{-j}$ is not known. In any case, the normal approximation
(to the binomial distribution) is asymptotic and not very accurate for
small $M$: therefore I propose using a 
jackknife test~\citep{efron79,efronandgong83}. I obtain a
jackknifed variance estimate of $\zeta_{-j}$ for  each
$j=1,2,\ldots,M$. Specifically, I calculate $\zeta_{-(j,k)} = \frac 2\pi
\arcsin\sqrt{{}_m^s\omega_{-(j,k)}} - \frac 2\pi
\arcsin\sqrt{{}_m^s\omega_{-k}}$, where  ${}_m^s\omega_{-(j,k)}$ is the
summarized overlap measure obtained using all but the $j$th and the
$k$th studies. The jackknifed variance estimator for $\zeta_{-j}$ is
then given by 
\begin{equation}
  {s}^2_{\zeta_{-j}} = \frac1{(M-1)(M-2)}\sum_{\substack{k=1 \\ k\neq 
      j}}^M
  \left[\zeta_{-(j,k)} - \bar\zeta_{-j}\right]^2,
\label{jack.variance}
\end{equation}
where  $\bar\zeta_{-j}$ is the jackknifed mean given by
\begin{equation*}
\bar\zeta_{-j} = \frac1{M-1} \sum_{\substack{k=1 \\ k\neq
  j}}^M\zeta_{-(j,k)}.
\end{equation*}
The test statistic for detecting significant reduction in the
summarized overlap measure upon including  the $j$th study 
(and hence detecting if it is anomalous) is given by 
\begin{equation*}
\tau_{-j} = \frac{\zeta_{-j}}{s_{\zeta_{-j}}},
\end{equation*}
with $p$-value computed from the area under the 
$t_{M-2}$ density to the right of
$\tau_{-j}$. False Discovery Rate~(FDR)-controlling techniques, such
as in~\citet{benjaminiandhochberg95} may be used to control for the
proportions of expected false discoveries in detecting significant
outliers. 

%I conclude discussion on identifying outlying and anomalous activation

Several comments are in order. First, I note that the multiple Jaccard
index proposed in~\citet{colwellandcoddington94} is unusable in our
testing strategy because its range of values changes from one set of
studies to the next. My summarized multiple Jaccard similarity
coefficient however takes values between zero and unity and is comparable
from one set of studies to the other: thus, it can be compared across
different jackknifed samples, which under the null hypothesis would have similar
distributional properties. Further, the outlier detection method
proposed in this section contrasts with 
that in \citet{woolrich08} in that the latter draws inferences on the
original post-processed time series data at each
voxel. ~\citet{mcnameeandlazar04} use maps of  $t$-statistics
or $p$-values of activation to identify 
outliers. My proposed approach uses downstream statistics in a similar
manner as the
latter, but only
requires activation maps which can be acquired by any method,
even methods that differ from one study to the next. In this regard,
it may be considered to be a generalization of~\citet{mcnameeandlazar04}'s
methodology.   

A reviewer has pointed out that the pairwise Jaccard similarity
coefficient  $_m\omega_{jl}$, while a
proportion, is still a ratio of the number of 
voxels activated in both the $j$th and the $l$th replications to the
number of voxels activated in at least one of them. This is also true
of the pairwise percent overlap measure of activation proposed
by~\citet{romboutsetal98} and  \citet{machielsenetal00}. 
The distribution
of the ratios of random variables is quite complicated and can bring
with it a host of issues~(see, for
instance, \citep{allisonetal95}). The use of the jackknife,
however means that no assumptions are made on the distribution of
either the numerator or the denominator in any of the pairwise
measures involved in the construction of the (jackknifed) variance
estimator in~(\ref{jack.variance}). This allays potential concerns
on distributional assumptions as a consequence of using ratios of
random variables. 
%Thus, concerns on distributional assumptions which govern the use of
%ratios of random variables is not an issue. 
\section{Illustration and Application to Motor-task Experiments}
\label{experiments}
\subsection{Experimental and Imaging Setup}
\label{imaging}
The methodology was applied to two replicated sets of experiments,
with each set corresponding to right- and left-hand finger-thumb opposition
tasks, performed by the same normal right-hand
dominant male volunteer, after obtaining his informed
consent. Each set of  experiments consisted of twelve sessions  
conducted over a two-month period. The experimental paradigm in
a session consisted of eight cycles of a simple finger-thumb
opposition motor task, with the experiment performed by the
right or left hand depending on whether the session was for the
right- or left-hand experiment. During each cycle, the subject
conducted finger-thumb opposition of the hand for 32 seconds, followed
by an equal period of rest.  
MR images for both experiments were acquired on a GE 1.5 Tesla Signa
system equipped  with echo-planar gradients, with 
inter-session differences minimized using~\citet{nolletal97}'s
recommendations on slice-positioning. 
 For
each fMRI session, a single-shot spiral sequence (TE/TR = 35/4000 ms)
was used to acquire twenty-four 6~mm-thick slices parallel 
to the AC-PC line and with no inter-slice gap. Thus, data were
collected at 128 time points.  Structural
$\mbox T_1$-weighted images were also acquired using a standard spin-echo
sequence~(TE/TR = 10/500 ms).
The data were
transferred from the scanner on to an SGI Origin 200 workstation where
image reconstructions were performed. Motion-related artifacts
in each replication were reduced via the Automated Image
Registration~(AIR) software using the default first image as target,
and then the time series at each 
voxel~\citet{woodetal98} was normalized to remove linear
drift. Further residual misregistration between the twelve sessions
was 
minimized by application of inter-session registration algorithms in
AFNI~~\citep{coxandhyde97}.  
Functional maps were created for each session after
computing voxel-wise $t$-statistics (and corresponding $p$-values) using a
general linear model, discarding the first three image volumes (to
account for $\mbox T_1$ saturation effects) and assuming first-order
autoregressive errors, using sinusoidal waveforms  with lags of 8 seconds.
The choice of waveform represented the BOLD response with the lag duration
corresponding to %the time the actual BOLD 
when the response was
seen after the theoretical start of the stimulus. Activation maps were
drawn using the {\tt R} package {\tt AnalyzeFMRI} and 
Random Field theory and the expected Euler
characteristic derivations of~\citet{adler81} and~\citet{worsley94} at a
significance level of 5\%. The methodologies
developed in this paper were applied to these maps. All computations
were written using the open-source statistical software~{\tt
  R}~\citep{R}.% and performed on a IBM T61 Thinkpad having a Intel(R)
              % Core(TM)2 Duo CPU     T7700  processor with
              % clock-speed 2.40GHz and  running the 64-bit Fedora 10
              % 2.6.27.24-170 Linux kernel

\subsection{Results}
\label{results}
\begin{figure*}[h]
\includegraphics[width=\linewidth]{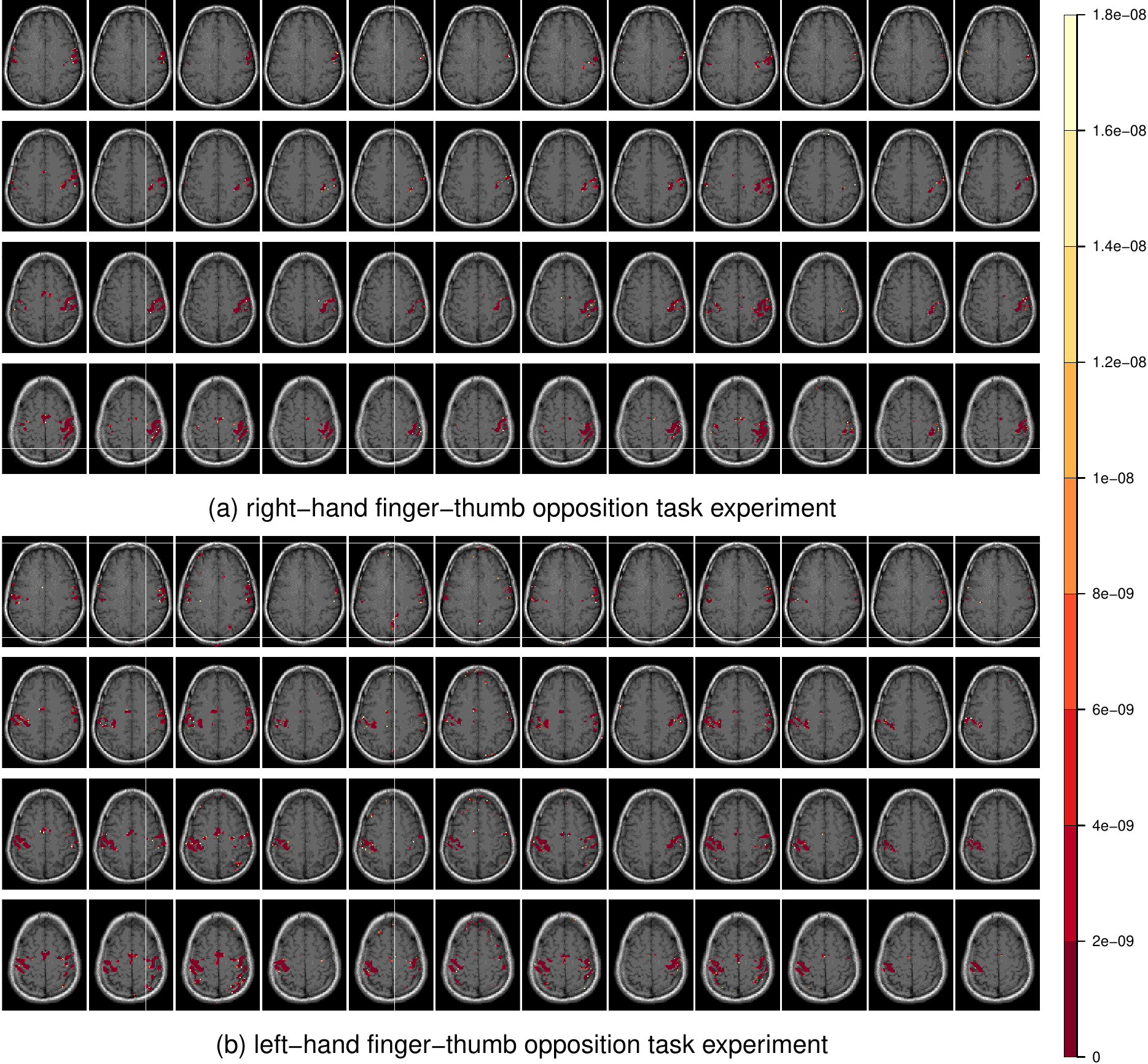} 
\caption{Radiologic view maps of observed $p$-values of activation of
  the $t$-test of motor function for (a) the right-hand and (b)
  left-hand finger-thumb opposition task experiments. For each set of
  experiments, we display radiologic view maps for the 18th, 19th,
  20th and 21st slices (row-wise). The twelve replications are
  represented column-wise from 1 through 12. 
  For each slice, we display the
  $p$-values of activation for the thresholded voxels using a $t$-test of
  the motor function for the twelve replications of the   finger-thumb
  opposition experiment performed by (a) the right hand and (b) the
  left hand of the same right-hand dominant male volunteer. Note the   differences in location
  and extent of activation over the twelve  replications. Note, also
  the substantial more variability in the experiments performed by the
  subject's left hand than on the right. 
 }
\label{abothhandrep}
\end{figure*}
Figure~\ref{abothhandrep} represents the observed $p$-values of
activation for slices 18, 19, 20 and 21 (row-wise) over the twelve
replications for both the (a) right-hand and (b) left-hand
finger-thumb opposition tasks. (All displayed maps in this paper  
are in radiologic views and overlaid on top of the 
corresponding $\mbox T_1$-weighted anatomical images.) The specific
slices were chosen for display because they encompass the ipsi- and
contra-lateral pre-motor cortices~(pre-M1), the primary motor 
cortex~(M1),  the pre-supplementary motor cortex (pre-SMA), and the
supplementary motor cortex~(SMA). Clearly, there is some variability
in the results for the right-hand task. In 
Figure~\ref{abothhandrep}a for instance, all experiments
identify activation in the  left M1 and in the ipsi-lateral pre-M1
areas, but there is some modest variability in the identified
activation in the contra-lateral pre-M1, pre-SMA and SMA voxels, with
some experiments %(most notably, the fifth, eleventh and to a lesser
                 %extent, twelfth replications) 
reporting very localized or no activation and others having these
regions as activated and somewhat diffused in extent. Slices for
the left-hand finger-thumb opposition task experiments in
Figure~\ref{abothhandrep}b, on the other hand, show far more
variability, both in location and extent. It is interesting to note
that while most experiments identify activation in the right M1, the
ipsi-lateral, contra-lateral pre-M1, pre-SMA and SMA areas, they also
often show activation in the corresponding left regions. The case of
the eighth replication is extremely peculiar. Most of the activity in
the four slices are in the left areas and the right areas have little
to no activation. This makes one wonder if
the naturally right-hand dominant male volunteer had, perhaps
unintentionally and out of habit, used his right hand instead of his
left in performing some part of the experimental paradigm. In summary,
there is clearly far more variability in the left hand set of
experiments  than in the right hand set. We now assess the reliability
in each set separately. 

\subsubsection{Reliability of right-hand finger-thumb   opposition
  task experiments} 
\label{righthand}
\begin{figure*}
\centerline
\mbox{
  \subfigure[$\Omega$]{\includegraphics[width=.5\linewidth]{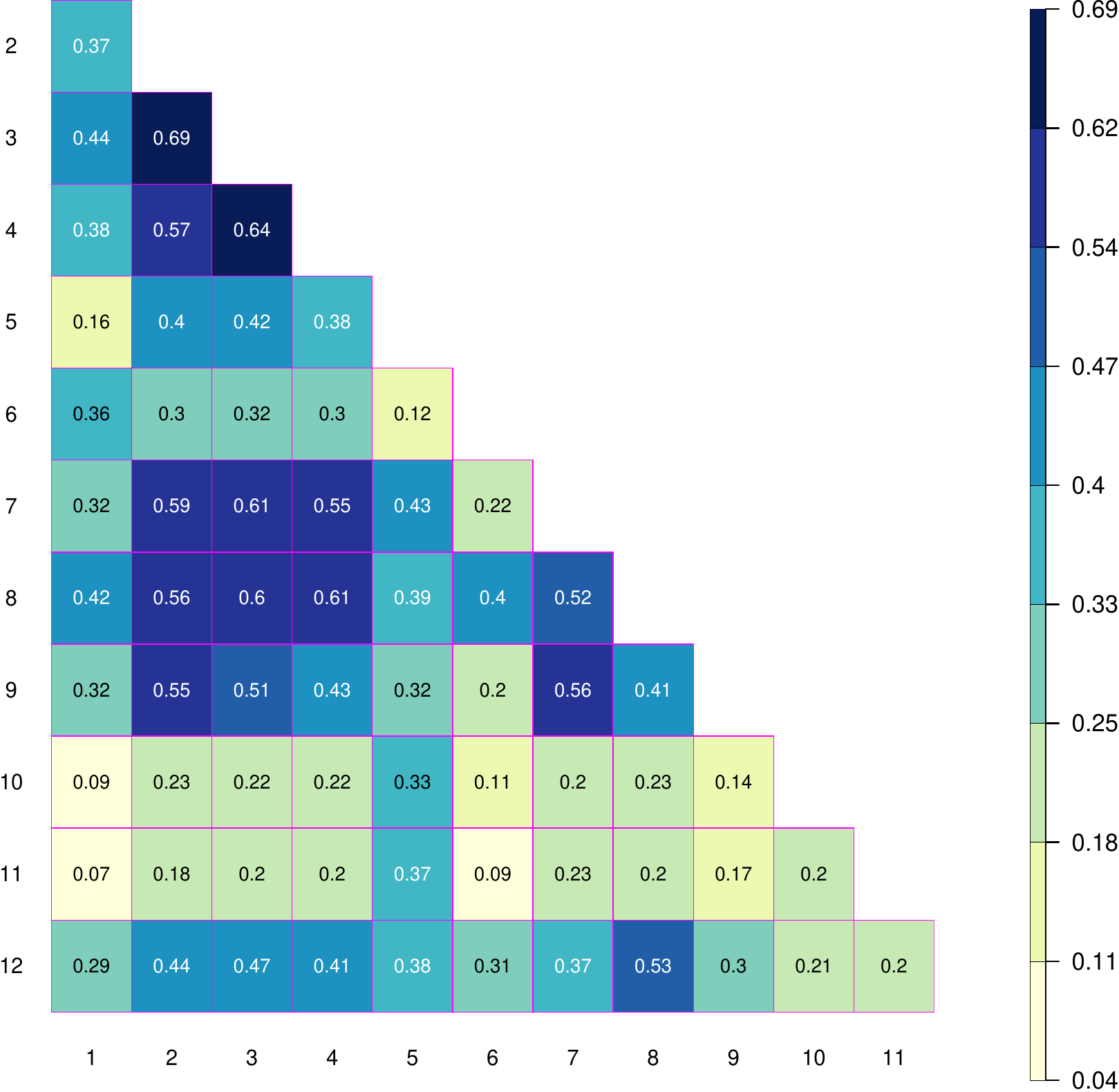}}
%}
%\mbox{
  \subfigure[${}_m\Omega$]{\includegraphics[width=.5\linewidth]{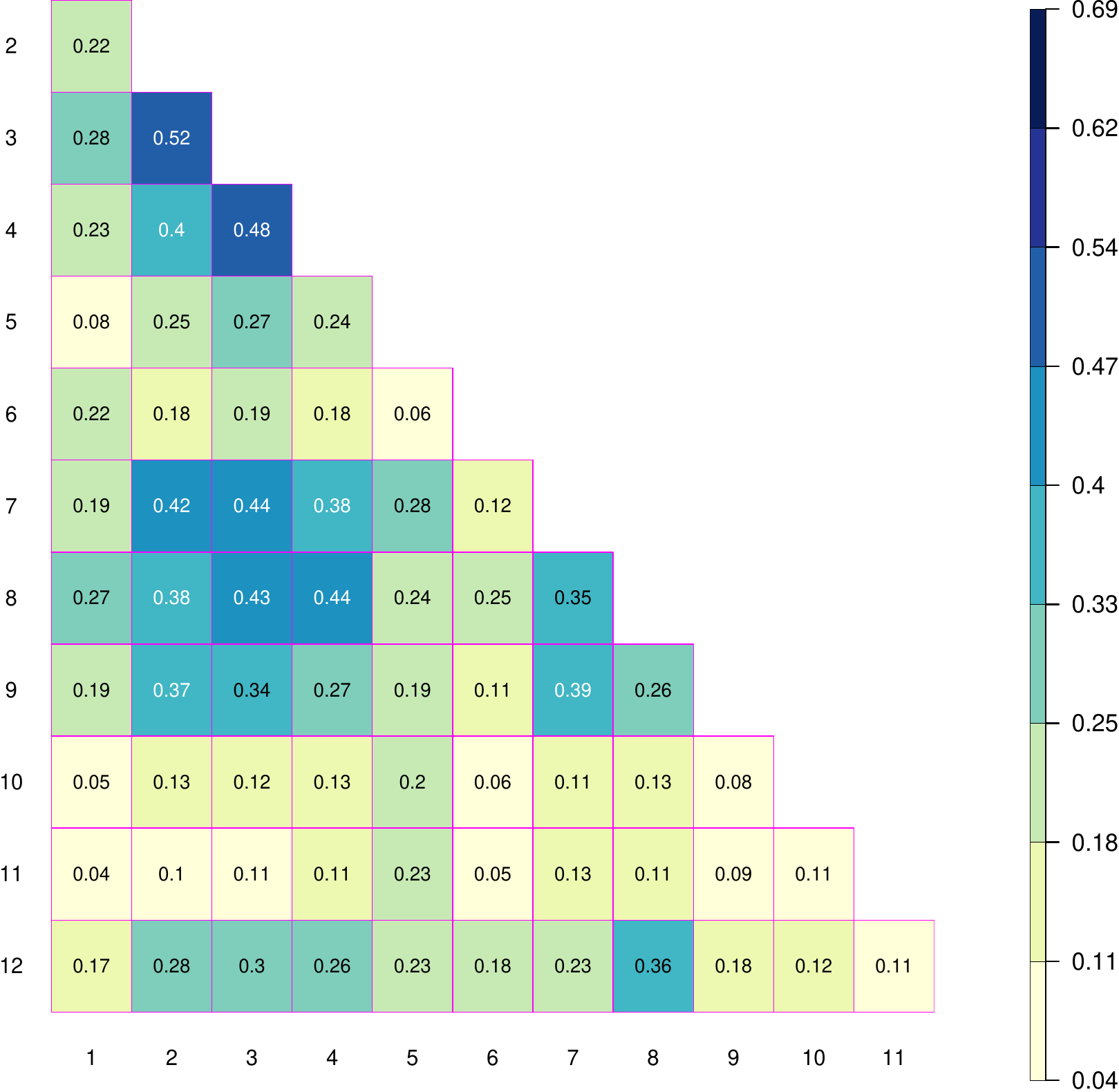}} 
}
\caption{Display of overlap measures on the right-hand
  finger-thumb-opposition task experiment, obtained using
  (a)~\citet{romboutsetal98}  
and~\citet{machielsenetal00}'s $\Omega$ and its (b) modified Jaccard
similarity coefficient version
$_m\Omega$ as proposed in this paper.}  
\label{rightoverlap}
\end{figure*}
Figure~\ref{rightoverlap} displays the lower triangle of the matrix
pairwise per cent overlap measures of activation calculated 
as per (a)~\citet{romboutsetal98} and~\citet{machielsenetal00} and (b)
the Jaccard similarity coefficient modification suggested in this
paper. %(Note that each map is displayed using its own palette.) 
As proved in Section~\ref{moa}, 
$_m\Omega \leq \Omega$, entry-wise. The (3,2) pair of experiments has
the highest percent overlap of activation, with $_m\omega_{3,2}= 0.524$
while $\omega_{3,2}= 0.688$. Compared to other values of $j$, the 
low $_m\omega_{j,l}$ values for $j=10$ or 11 are striking! While
$\omega_{j,l}$s are lower for $j=10,11$ than for other $j$, they are
not as pronounced as $_m\omega_{j,l}$. 
This points to the possibility that activation maps obtained using
these replications may be somewhat different from the 
others. Using the methodology of Section~\ref{summary} provides us
with the summarized measures $_m^s\omega =  0.244$ and $^s\omega =
0.372$. We now investigate performance of the  methodology of
Section~\ref{outliers} in flagging potential outliers or anomalous
studies.  
                                                              
\begin{figure}
\includegraphics[width=\linewidth]{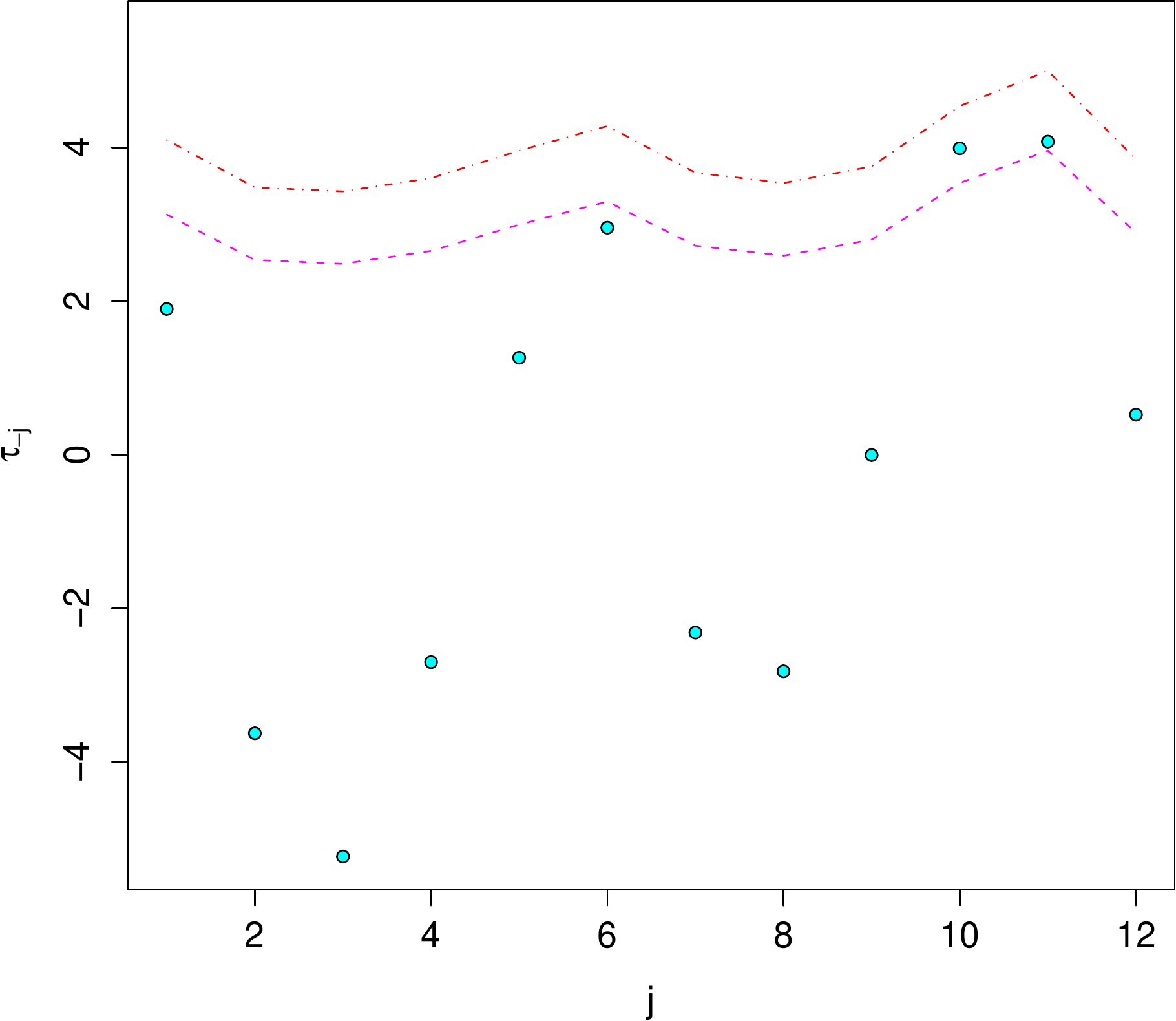}
\caption{Plot of $\tau_{-j}$ against $j$ for the right-hand 
  finger-thumb opposition task experiment. The half-broken  and
  semi-solid lines
  displays thresholds obtained when controlling eFDR at $q=0.05$ 
  and $q=0.01$, respectively.}
\label{righthandoutliers}
\end{figure}
\begin{figure*}
\centering
\mbox{
  \hspace{-0.3in}
  \subfigure[]{\includegraphics[width=.5\linewidth]{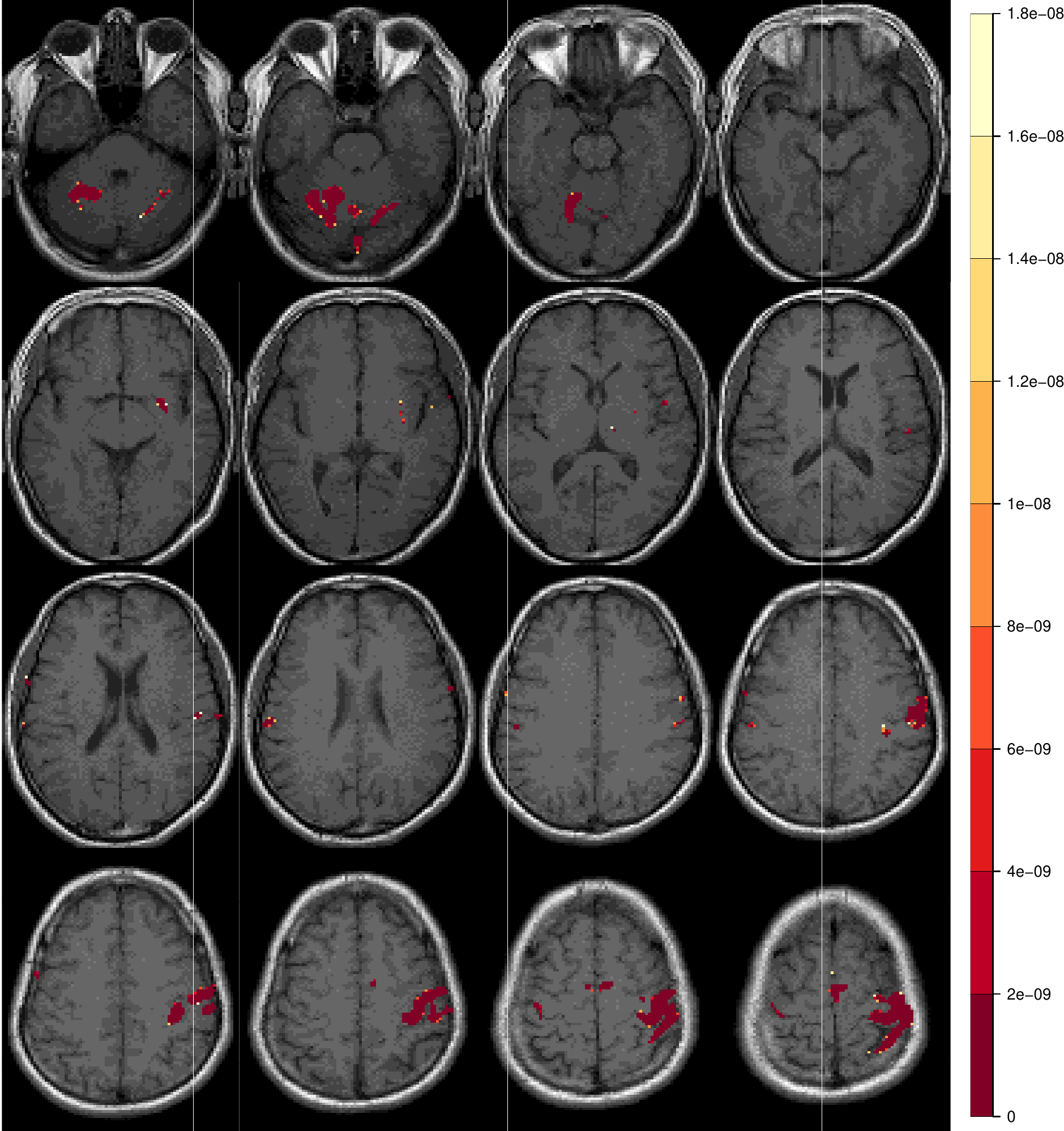}} 
  \subfigure[]{\includegraphics[width=.5\linewidth]{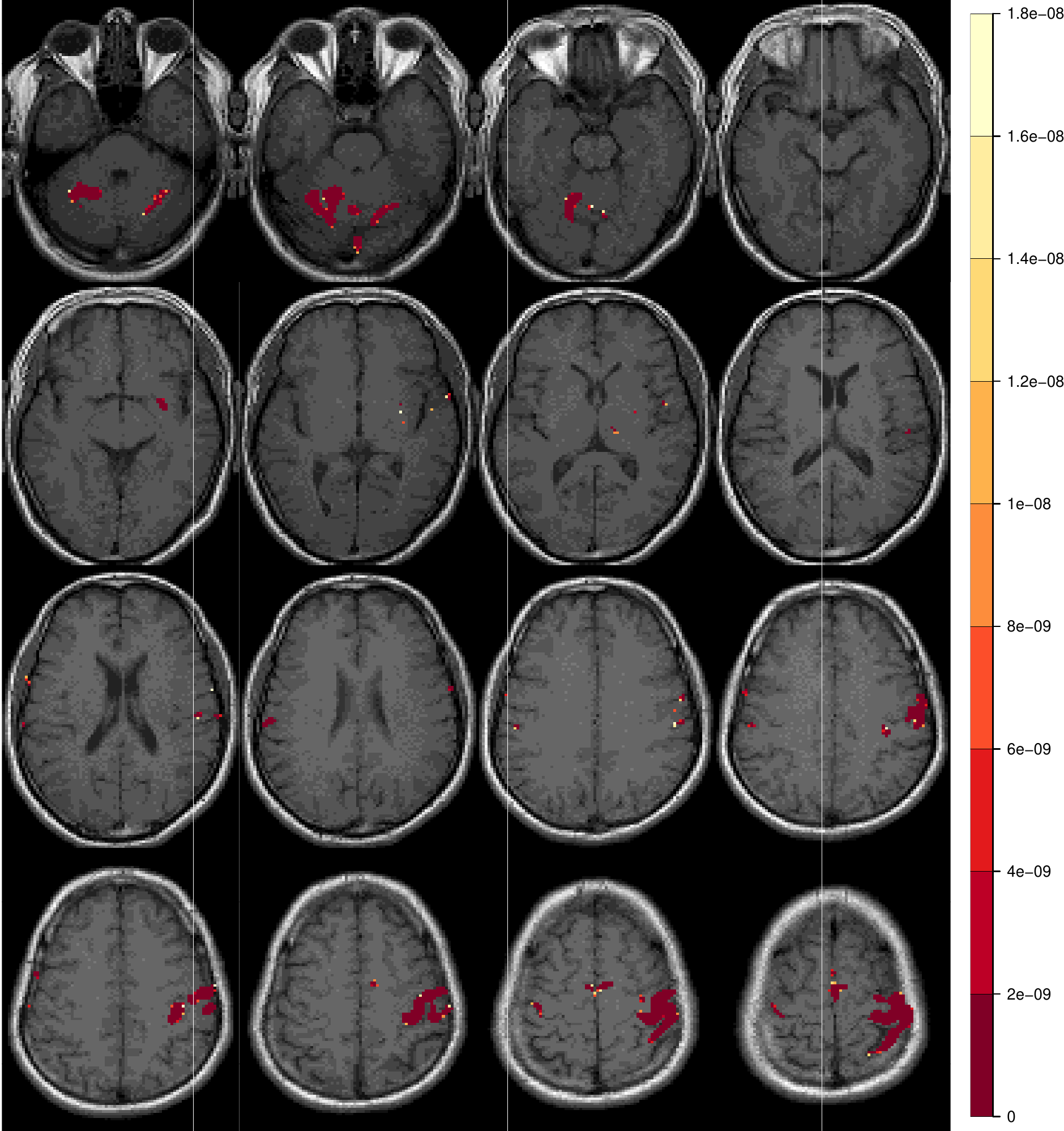}}  
}
% \mbox{
%\hspace{-0.2in}
% \subfigure[]{\includegraphics[width=2.5in]{righthandmodreddiff}}  
%}
%\includegraphics[height = 4in, width=3.75in]{righthandallfecrel}

\caption{Composite activation map for slices 7 through 22, of the
  right-hand finger-thumb opposition task experiment obtained using 
  (a) all twelve fMRI studies  and (b) all but the tenth and eleventh 
  studies. Displays are as in Figure~\ref{abothhandrep}.}
\label{righthandallfecrel}
\end{figure*}

The coefficient of variation in the jackknife-estimated standard
deviations of $\hat\zeta_{-j}$s was
around 0.0501: this indicates that the arc-sine transformation was
able to stabilize the variance substantially. 
Figure~\ref{righthandoutliers} plots the computed $\tau_{-j}$ against 
$j$ for each $j=1,2,\ldots,12$. Note that the values of $\tau_{-10}$
and $\tau_{-11}$ are fairly high: indeed, the corresponding fMRI maps
would be identified as significant outliers if we used an expected
FDR~(eFDR) of
$q=0.05$, but not so using an eFDR of $q=0.01$. Thus they may be
considered to be {\em moderate outliers}: this finding
is in keeping  with the general impression we obtained from
Figure~\ref{abothhandrep}a. Eliminating the moderate outliers
increases the summarized overlap measure: ${}^s_m\omega_{-(10,11)} =
0.287$ (${}^s\omega_{-(10,11)} =
0.432$). Figure~\ref{righthandallfecrel}a and b displays the composite
activation map obtained upon combining all the replications and all
but the tenth and eleventh replications, respectively. Each composite
map was obtained by averaging the $t$-statistics for each
study~\citep{lazaretal02} and determining activation as before using
the Random Field theory of~\citet{worsley94} at 5\% significance.
The activated regions in Figure~\ref{righthandallfecrel}b are slightly
more defined than in Figure~\ref{righthandallfecrel}a. This makes
sense because the effects of the less reliable studies have been
removed in constructing the composite activation map of
Figure~\ref{righthandallfecrel}b. 

\subsubsection{Reliability of left-hand finger-thumb   opposition
  task experiments} 
\label{lefthand}
\begin{figure*}
\centerline
\mbox{
  \subfigure[$\Omega$]{\includegraphics[width=.5\linewidth]{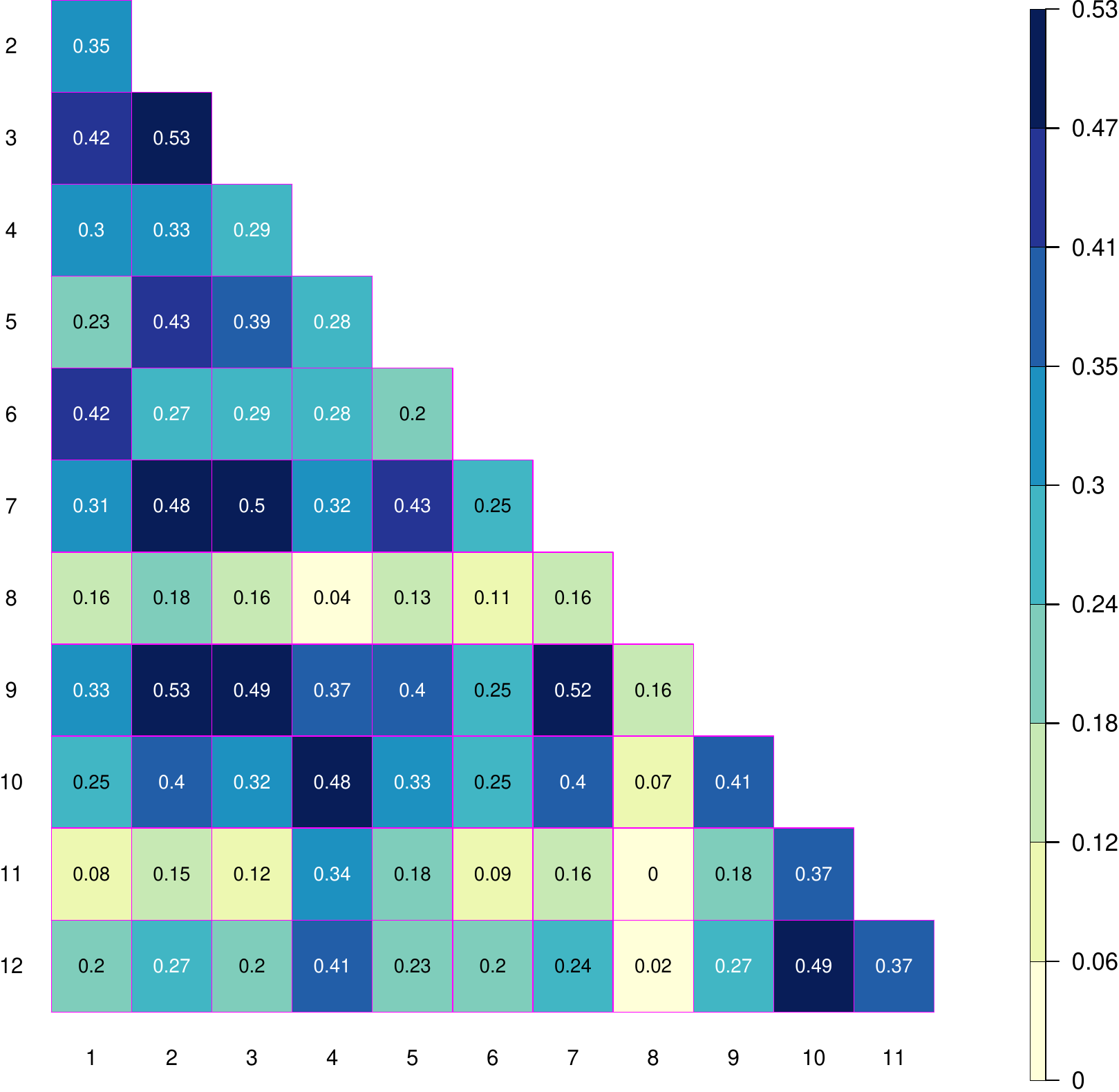}}
%}
%\mbox{
  \subfigure[${}_m\Omega$]{\includegraphics[width=.5\linewidth]{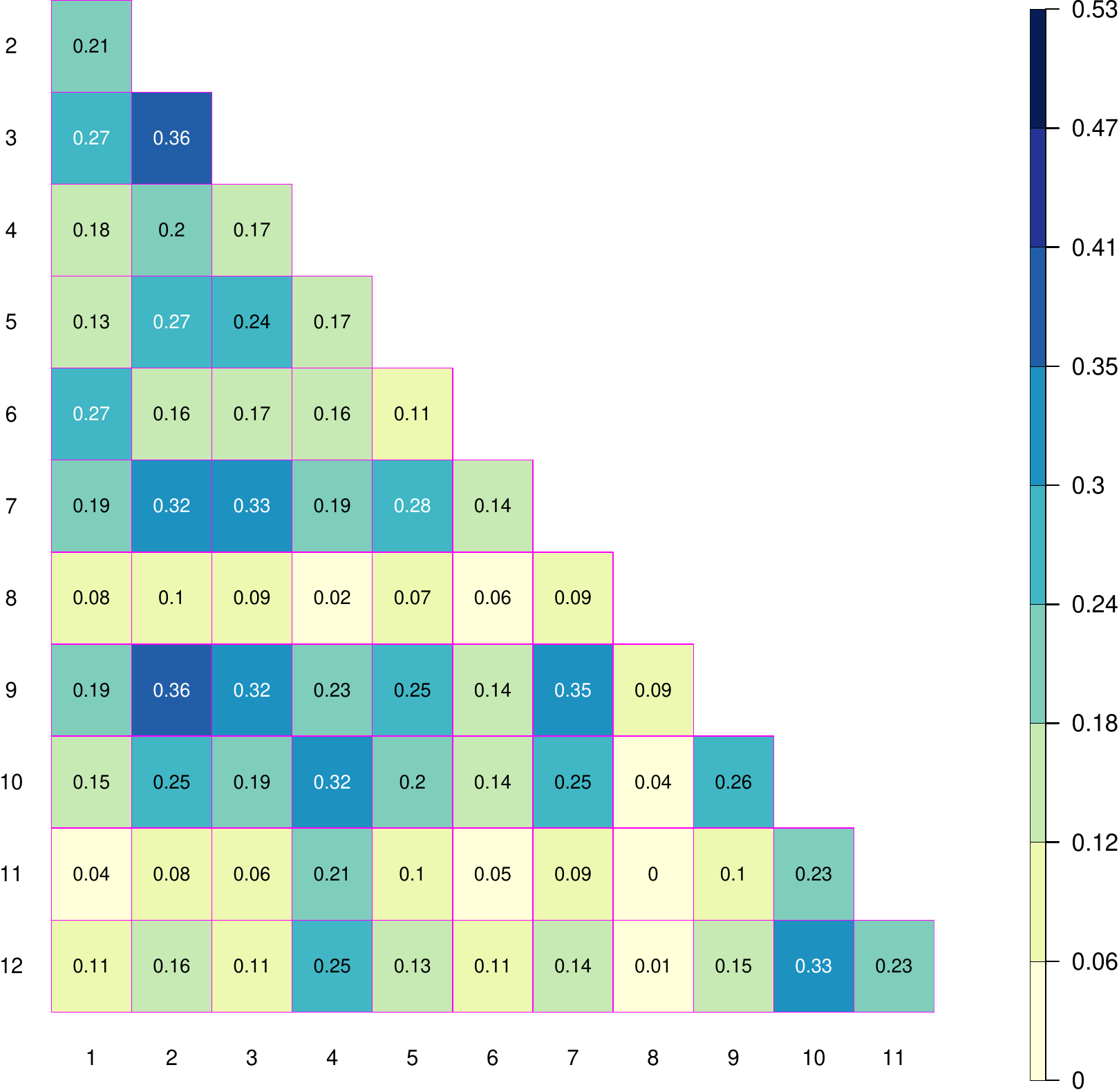}} 
}
\caption{Display of overlap measures on the left-hand
  finger-thumb-opposition task experiment, obtained using
  (a)~\citet{romboutsetal98}  
and~\citet{machielsenetal00}'s percent overlap measures of activated
and their (b) modified (Jaccard similarity coefficient) versions as
proposed in this paper. Displays 
are analogous to Figure~\ref{rightoverlap}.} 
\label{leftoverlap}
\end{figure*}
Figure~\ref{leftoverlap} displays the lower triangle of $\Omega$ and
${}_m\Omega$ for this set of experiments. 
Once again, $_m\Omega \leq \Omega$ element-wise. The second and the ninth
replications have the highest  percent overlap of activation 
($_m\omega_{9,2} = 0.361$ and $\omega_{9,2} = 0.531$) while there is
hardly any overlap between the eighth and the 
eleventh replications ($_m\omega_{11,8} = 0.002$, $\omega_{11,8} =
0.004$). In general, the values of $_m\omega_{8,j}$ (and $\omega_{8,j}$)
are very low for all $j\neq 8$ and in line with our suspicions from
studying Figure~\ref{abothhandrep}b. The graphical representation of 
$\omega_{8,2}$ and $_m\omega_{8,2}$ in Figure~\ref{leftoverlap}
presents the case for using ${}_m\omega$ over $\omega$ very well. In
Figure~\ref{leftoverlap}a, the value of $\omega_{8,2}$ is in the
middle third of the scale for $\Omega$: thus, the graphical display would
cause us to hesitate before 
declaring that the activation identified in the eighth and second
replications are very different from each other. However,
Figure~\ref{abothhandrep}b does not provide much justification for
such second thoughts, corroborating the value of $_m\omega{8,2}$ (which is
in the lower third of the values graphically displayed in
Figure~\ref{leftoverlap}b). Values of $_m\omega_{11,j}$ (and
$\omega_{11,j}$) are also low for all $j\neq 
11$ even though they are a bit higher than for $_m\omega_{8,j}$ and
$\omega_{8,j}$. The summarized measure over all twelve replications
was $^s_m\omega = 0.187$ and $^s\omega = 
0.303$. Thus, there is far less reliability in identified activation
in this set of experiments relative to the right-hand set. We now
identify potentially anomalous fMRI studies in the left-hand set.  

\begin{figure}
\includegraphics[width = \linewidth]{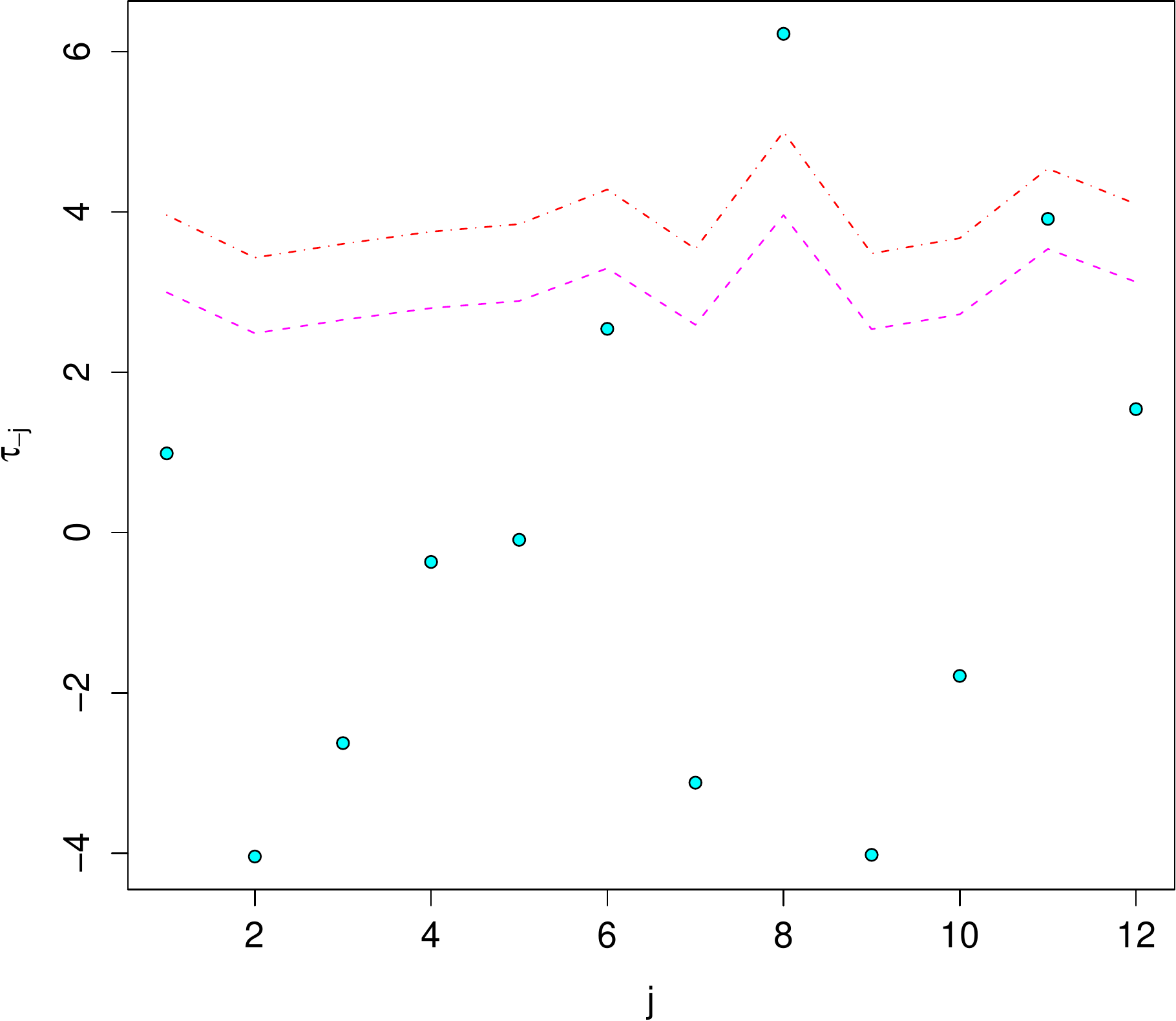}
\caption{ Plot of $\tau_{-j}$ against $j$ for the left-hand
  finger-thumb opposition task experiment, with displays as in 
  Figure~\ref{righthandoutliers}.}
\label{lefthandoutliers}
\end{figure}

Once again, the coefficient of variation in the jackknife-estimated
standard deviations of $\zeta_{-j}$ is small, around 0.093 so that the
variance stabilizing transformation is seen to do a good job in terms
of homogenizing variances in both sets of our experiments. 
Figure~\ref{lefthandoutliers} plots $\tau_{-j}$ against
$j$. Note that $\tau_{-8}$ is significant even when controlling eFDR
at $q=0.01$. The eighth study is thus an {\em extreme
  outlier}. The eleventh replication is, however, a {\em moderate
  outlier}. Deleting the eighth replication yields $^s_m\omega_{-8} =
0.204$ ($^s\omega_{-8}=0.329$) while deleting the moderate  and extremely
anomalous  studies results in  $^s_m\omega_{-(8,11)} = 
0.219$ ($^s\omega_{-(8,11)}=0.351$). 

\begin{figure*}
  \centering
  \mbox{
  \subfigure[]{\includegraphics[width=.5\textwidth]{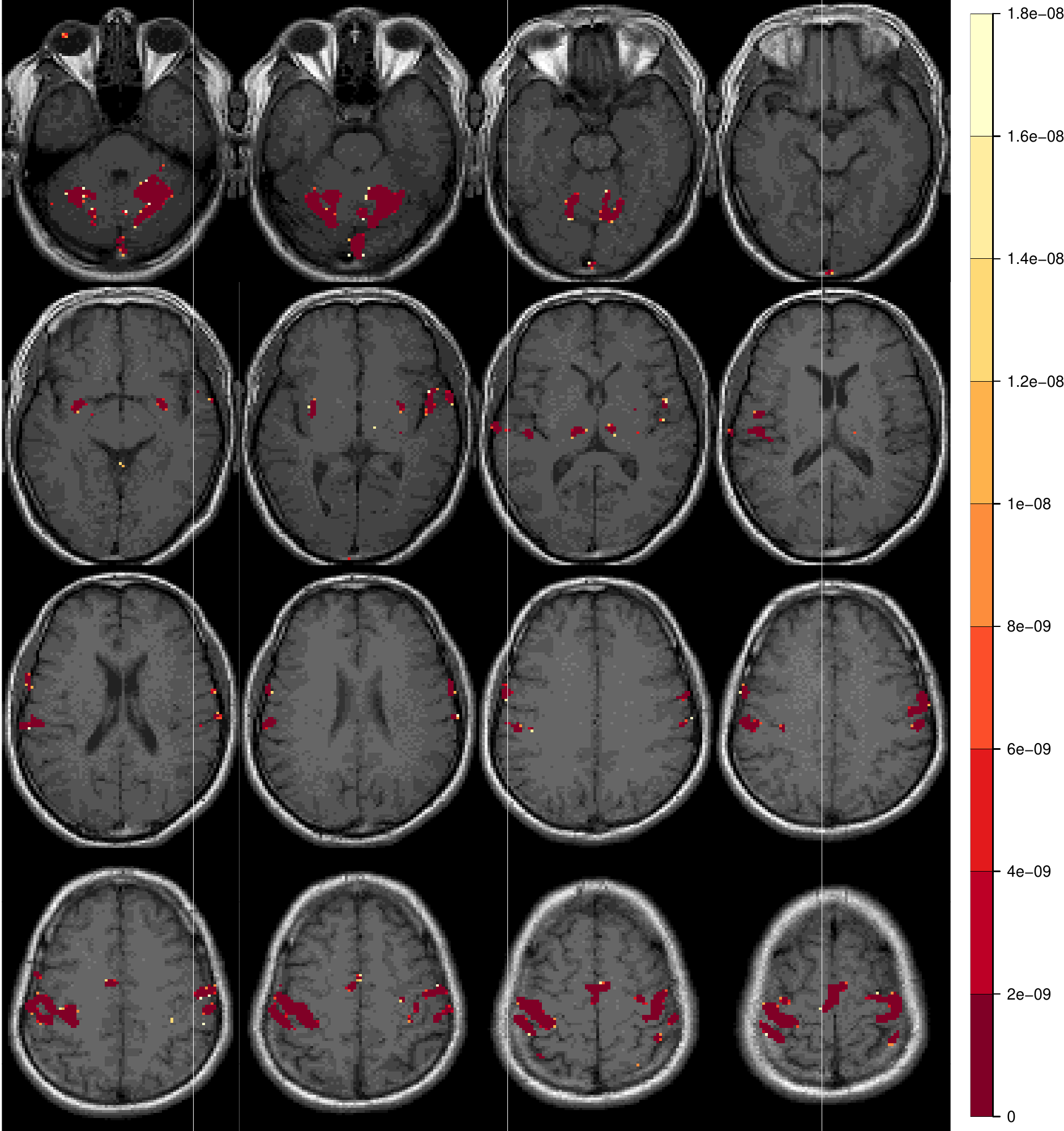}}
  \subfigure[]{\includegraphics[width=.5\textwidth]{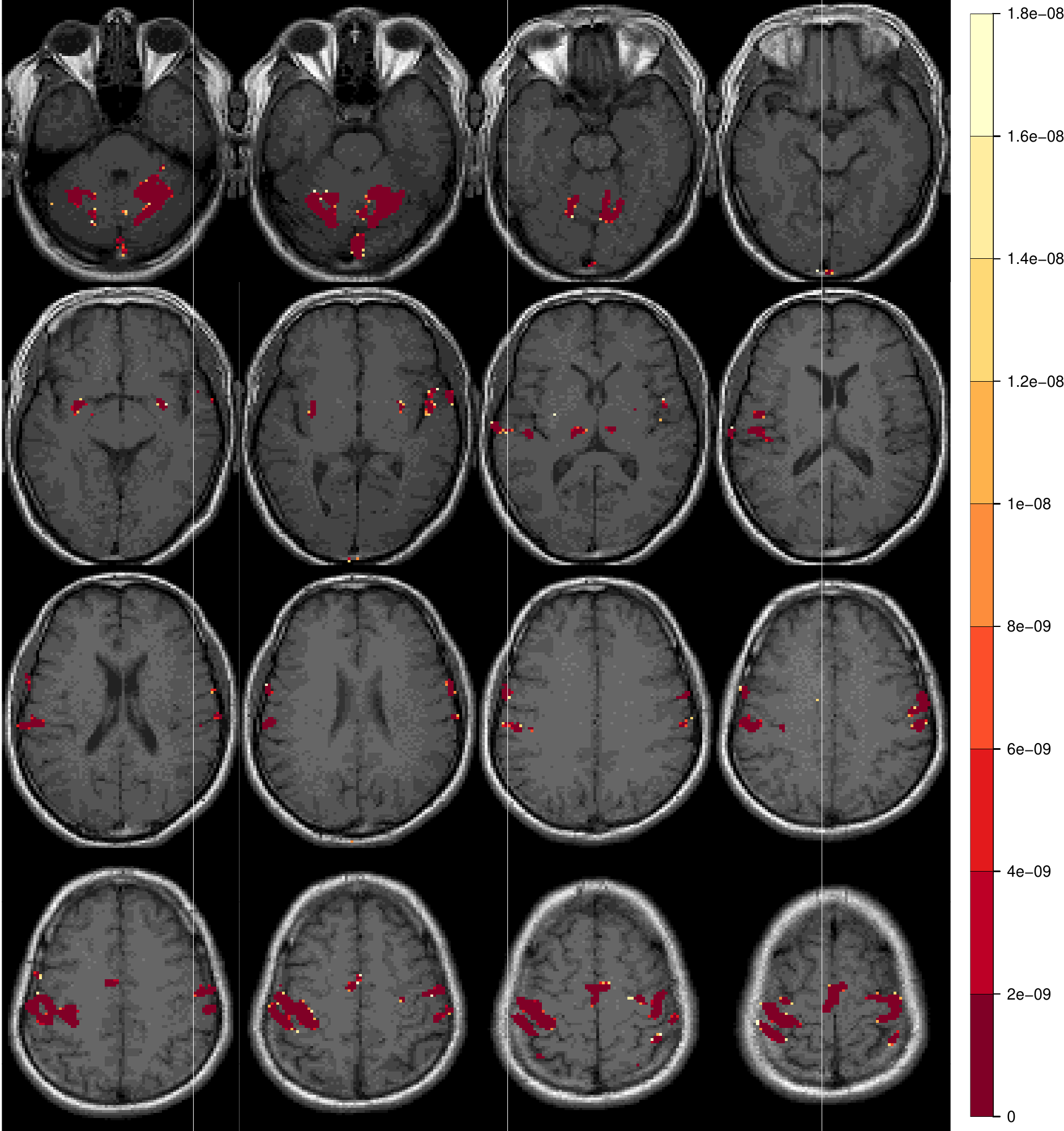}}}
\mbox{
  \subfigure[]{\includegraphics[width=.5\textwidth]{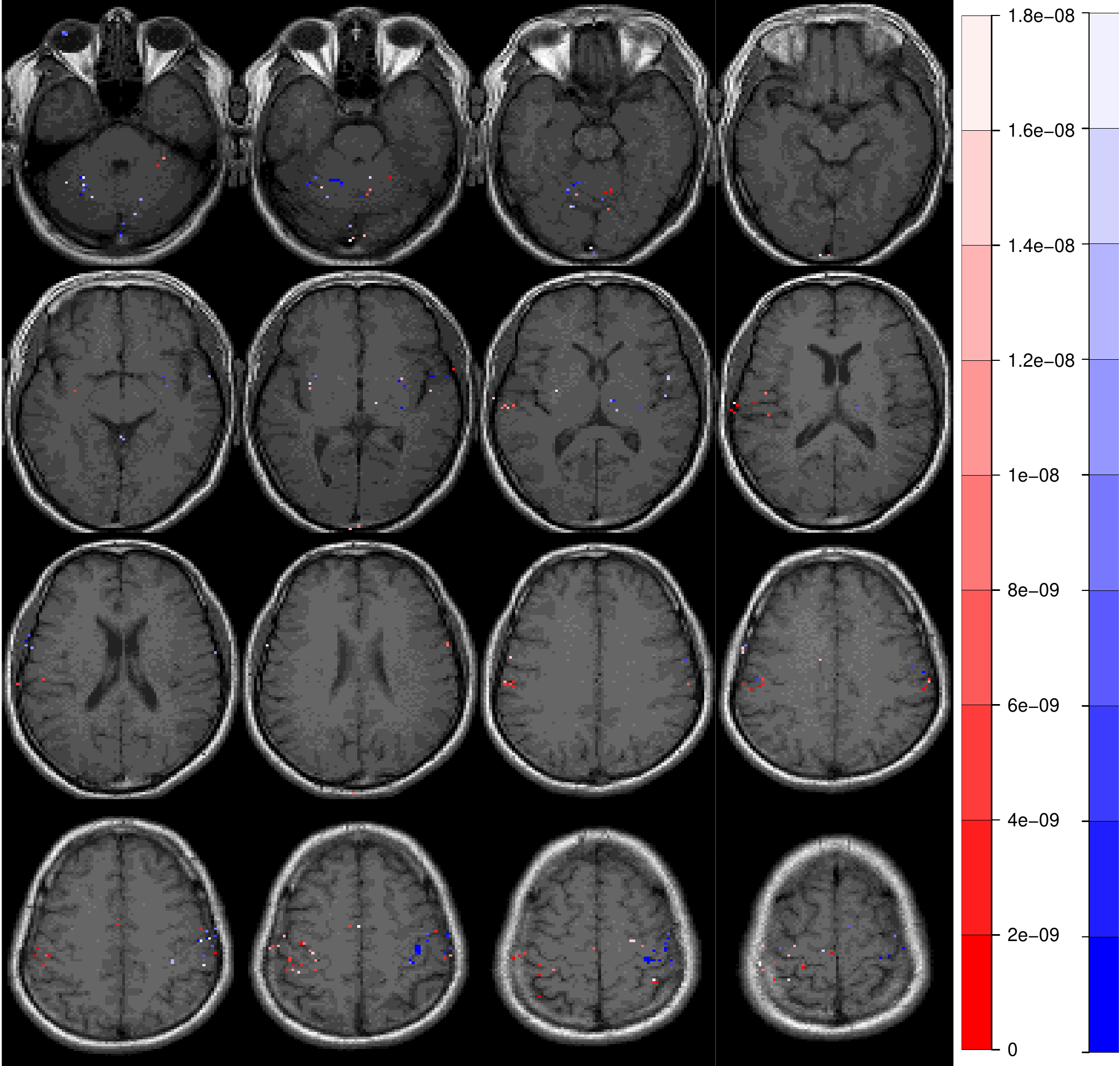}}}
%  \subfigure[]{\includegraphics[width=3.5in]{lefthandmodredfecrel}}

%\mbox{
%  \hspace{-0.3in}
%  \subfigure[]{\includegraphics[width=2.5in]{lefthandextreddiff}}
%  \hspace{-0.2in}
%  \subfigure[]{\includegraphics[width=2.5in]{lefthandextmodreddiff}}
% \hspace{-0.2in}
%  \subfigure[]{\includegraphics[width=2.5in]{lefthandextmoddiff}}
%%  \subfigure[]{\includegraphics[width=3.5in]{lefthandmodredfecrel}}
%}
\caption{Composite activation maps obtained using (a) all 
  studies and (b) all but the eighth fMRI study for the left-hand
  finger-thumb opposition task experiment. Displays are as in
  Figure~\ref{abothhandrep}. (c) Difference in the 
  composite activation maps of (a) and (b). Blue-hued voxels are those 
  that were identified as activated in (a) but not in (b) while
  red-hued voxels are those identified as activated in (b) but not in
  (a). In both cases, hue is proportional to the $p$-value of the test
  statistic when it was identified as significant. 
}
\label{lefthandcomposites}
\end{figure*}
Figures~\ref{lefthandcomposites}a and b display the composite
activation maps for the left-hand set by combining all studies and all
but the eighth studies, respectively. Figure~\ref{lefthandcomposites}c
shows voxels that were differentially activated in the two composite
maps. Slices 18 through 21 have more significant voxels in the left 
areas of (a) than in (b) and fewer significant voxels in the right
areas of (a) than in (b). There is therefore increased localization in the
identified activation when the eighth study is 
excluded.   
\subsubsection{Sensitivity to thresholding values}
\label{sensitivity}
\begin{figure*}
  \hspace{-0.1in}
\mbox{
  \subfigure[right-hand experiments, $\alpha=0.01$]{\includegraphics[width=.5\linewidth]{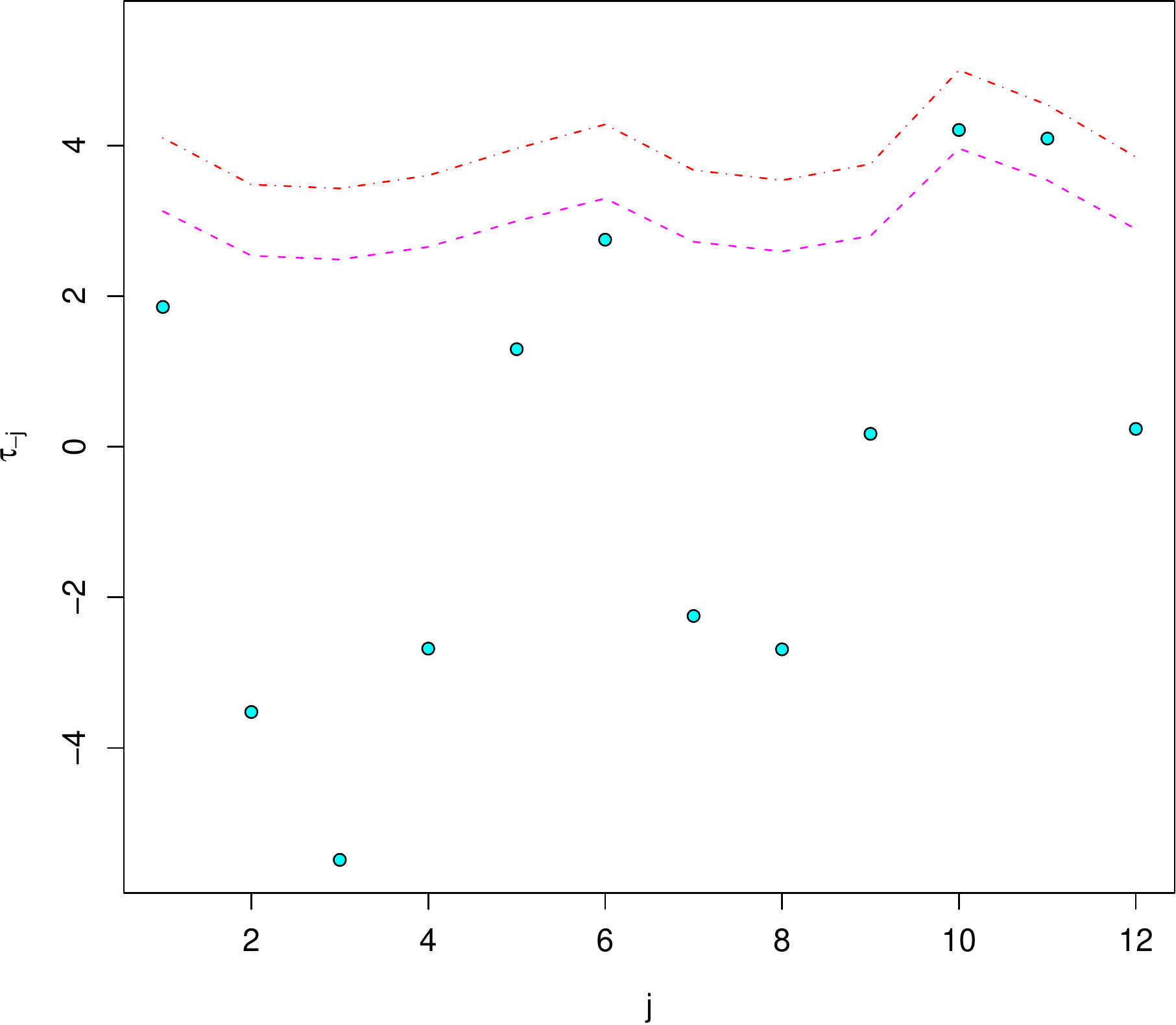}}
  \subfigure[right-hand experiments, $\alpha = 0.001$]{\includegraphics[width=.5\linewidth]{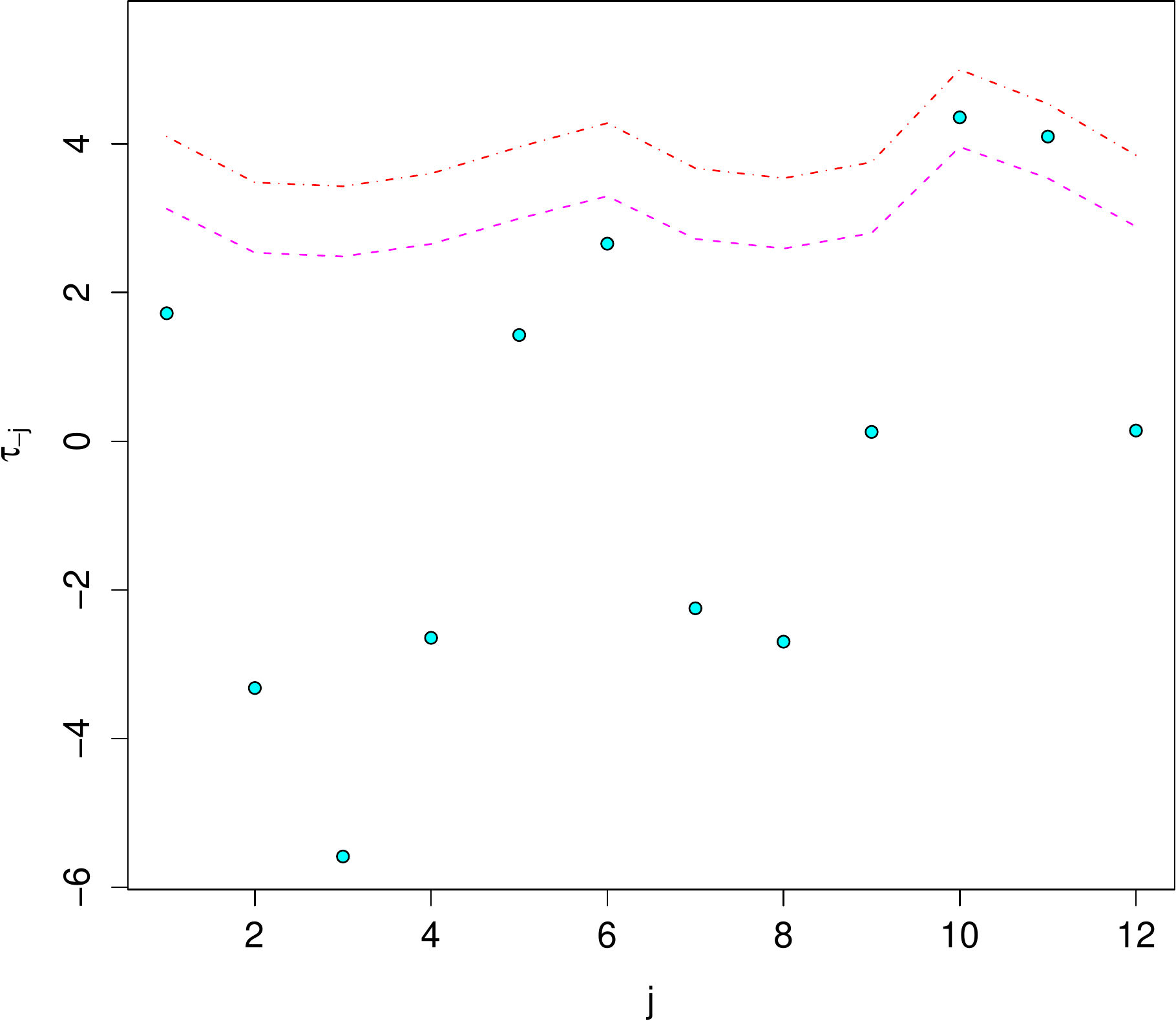}}
}
\mbox{
  \subfigure[left-hand experiments, $\alpha=0.01$]{\includegraphics[width=.5\linewidth]{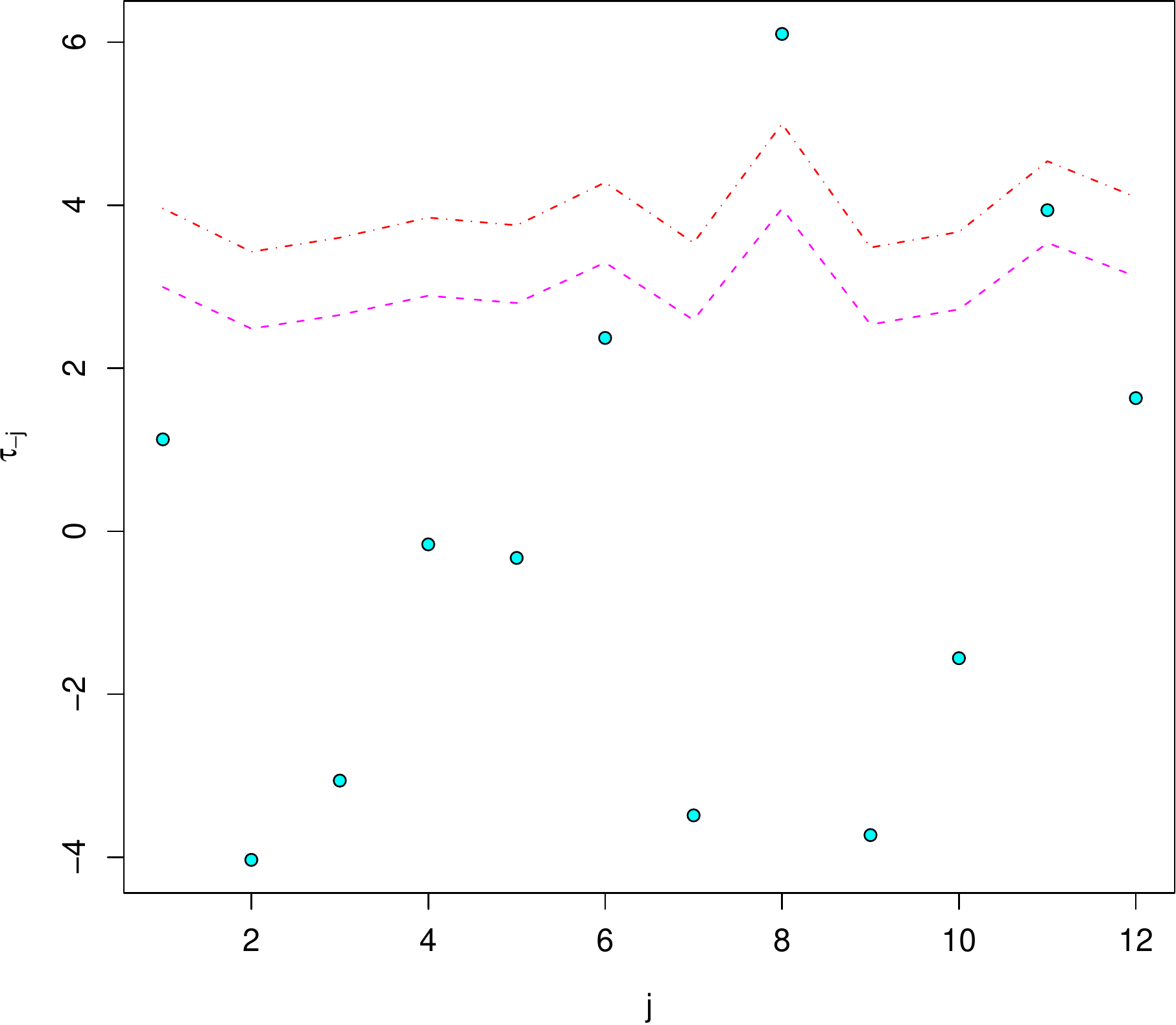}}
  \subfigure[left-hand experiments, $\alpha=0.001$]{\includegraphics[width=.5\linewidth]{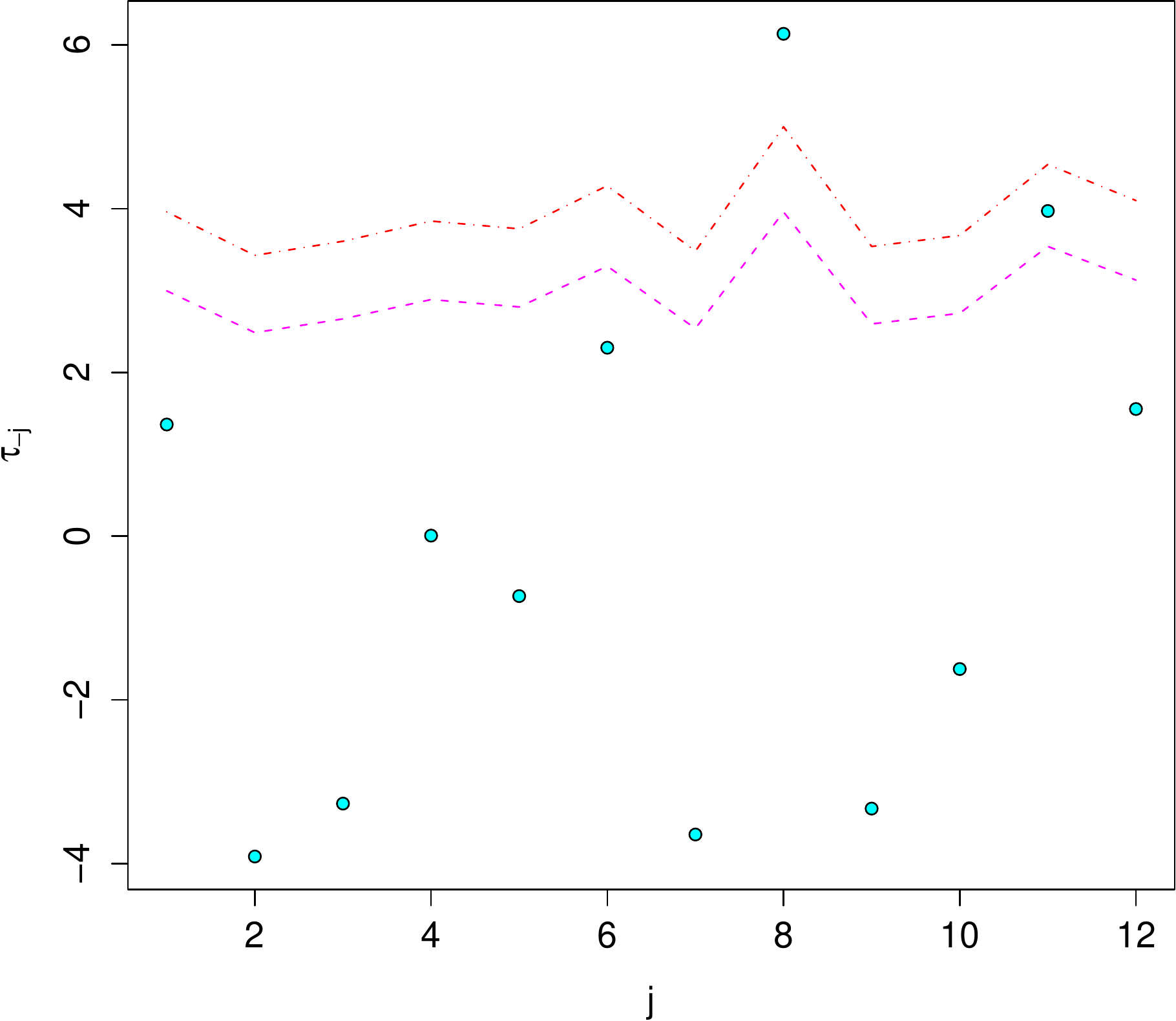}}
}
\caption{Plot of $\tau_{-j}$ against $j$ at thresholding values of (a,
  c) $\alpha = 0.01$ and (b, d) $\alpha = 0.001$ for (a, b) the right-
  and (c, d) left-hand finger-thumb opposition task
  experiments. Displays are as in Figure~\ref{righthandoutliers}.
}
\label{robustoutliers}
\end{figure*}
A reviewer has very kindly pointed out that the outliers identified in
Sections~\ref{righthand} and~\ref{lefthand} are from fMRI activation 
maps drawn using Random Field theory and at a significance threshold
of $\alpha=0.05$. The 
methodology was therefore applied to activation maps drawn using the
same approach but with more conservative significance thresholding
($\alpha=0.01$ and $\alpha=0.001$). Figure~\ref{robustoutliers}
displays the $\tau_{-j}$'s 
obtained from activation maps at the two significance thresholds
for the right- and the left-hand experiments. The tenth and eleventh
fMRI studies are again the only ones identified as moderately
anomalous for the right-hand experiments at both $\alpha=0.01$ and
$\alpha=0.001$. As before, the eighth fMRI study is also the only
extreme outlier in the left-hand set while the eleventh
study is the only moderate outlier. The outlier-detection
strategy thus appears to be remarkably robust to the exact significance
thresholding selected in creating our fMRI maps. 

In this section, I have demonstrated use of the Jaccard similarity
coefficient as a modified measure for the pairwise percent overlap of
activation and shown that it  
can provide a better sense of reliability. I have also illustrated
the use of my  summary measure for quantifying the 
overall percent overlap of activation from multiple fMRI
studies. Finally, I have illustrated the utility of the developed
testing tool to identify potentially anomalous fMRI maps and have also
shown that it is fairly robust to different choices of thresholding
used  in the preparation of fMRI activation maps.

\section{Discussion}
\label{discussion}
~\citet{romboutsetal98} and~\citet{machielsenetal00} have proposed a
measure of the percent overlap in voxels that are identified as
activated in any pair of replications. Although novel in the context
of studying fMRI reproducibility, this measure is the same as the Dice
coefficient~\citep{dice45,sorensen48} which is known to have several
drawbacks. This paper has investigated use of the Jaccard similarity
coefficient by slightly modifying the ~\citet{romboutsetal98}
and~\citet{machielsenetal00} measure. The modified measure is seen to
incorporate a more intuitive set theoretic interpretation, which is 
demonstrated  through some illustrative examples as well as through
application to two replicated fMRI datasets. A
summarized percent overlap measure of activation~(the summarized
multiple Jaccard similarity coefficient)  for quantifying
reliability of activation over multiple fMRI
studies has also been
proposed. A testing strategy has also been developed that uses
improvements in the summarized multiple Jaccard similarity coefficient 
upon excluding studies to evaluate whether a particular study is an
outlier or an anomaly and should be discarded from the
analysis. Although developed and  demonstrated  on test-retest
studies with replicated activation maps on a single subject, the
methodology is general enough to apply to multi-subject fMRI data. 

We have applied our developed methodologies to two sets of replicated 
experiments performed by the same right-hand-dominant normal male
volunteer. The two sets of experiments pertained to finger-thumb
opposition tasks performed by the subject using his right and his left
hand respectively. Our summarized measures of percent overlap of
activation are substantially higher for the right-hand task than for
the left-hand task. This agrees with the visual cues  provided in Figure~\ref{abothhandrep} where we 
noticed substantially more variability in the activation
maps for the left-hand task experiments than for the right-hand
ones. We have further used our testing strategy to flag down potentially
anomalous replications: for the right-hand task experiment, there were
two moderately anomalous studies. For the set of experiments on the
left-hand tasks, the eighth replication was an extreme anomaly while
the eleventh study was moderately anomalous. Deleting these studies
resulted in both increased localization and spatial extent of the
composite fMRI maps. Finally, the outlier detection was seen to be
remarkably insensitive to the choice of thresholding used in the
creation of the original activation maps.

There are a number of benefits that our suggested testing mechanism
for flagging anomalous studies provides. A study identified as a
potential outlier may trigger further investigation since there 
are several reasons why a study may be flagged as anomalous. For one,
it may point to physical issues with 
regard to the scanner. Alternatively, and in the context of
multi-subject studies, this may be useful for clinical
diagnosis: for example, it may 
be worth investigating why a particular subject had a very different
activation map. In other words, this can point the researcher  and the
neurologist to the need for further clinical  investigation and diagnosis. In
the testing scheme developed in Section~\ref{outliers}, we only
evaluated the effect of removing one 
observation at a time. It would be interesting to investigate the
effect of removing multiple observations. Another avenue worth
pursuing is the development of similar measures for grouped
fMRI studies. Thus, while this paper has made a promising
contribution, several issues meriting further attention remain. 

\section{Acknowledgments}
\label{acknowledgments}
I am very grateful to the Section Editor, the Handling Editor and
three reviewers whose very detailed and 
insightful comments on an earlier version of this manuscript greatly
improved its quality. I thank Rao P. Gullapalli of the University of
Maryland School of 
Medicine for providing me with the data used in this study.  This material is based,
  in part, upon work supported by the National Science
  Foundation~(NSF) under   its CAREER Grant No.  DMS-0437555 and 
by the National Institutes of Health~(NIH) under its Grant
No. DC-0006740.  

%\thebibliographystyle{elsarticle-harv}
\bibliographystyle{IEEEtran}
\bibliography{rm}

\begin{thebibliography}{10}
\providecommand{\url}[1]{#1}
\csname url@samestyle\endcsname
\providecommand{\newblock}{\relax}
\providecommand{\bibinfo}[2]{#2}
\providecommand{\BIBentrySTDinterwordspacing}{\spaceskip=0pt\relax}
\providecommand{\BIBentryALTinterwordstretchfactor}{4}
\providecommand{\BIBentryALTinterwordspacing}{\spaceskip=\fontdimen2\font plus
\BIBentryALTinterwordstretchfactor\fontdimen3\font minus
  \fontdimen4\font\relax}
\providecommand{\BIBforeignlanguage}[2]{{%
\expandafter\ifx\csname l@#1\endcsname\relax
\typeout{** WARNING: IEEEtran.bst: No hyphenation pattern has been}%
\typeout{** loaded for the language `#1'. Using the pattern for}%
\typeout{** the default language instead.}%
\else
\language=\csname l@#1\endcsname
\fi
#2}}
\providecommand{\BIBdecl}{\relax}
\BIBdecl

\bibitem{romboutsetal98}
S.~A. Rombouts, F.~Barkhof, F.~G. Hoogenraad, M.~Sprenger, and P.~Scheltens,
  ``Within-subject reproducibility of visual activation patterns with
  functional magnetic resonance imaging using multislice echo planar imaging,''
  \emph{Magnetic Resonance Imaging}, vol.~16, pp. 105--113, 1998.

\bibitem{machielsenetal00}
W.~C. Machielsen, S.~A. Rombouts, F.~Barkhof, P.~Scheltens, and M.~P. Witter,
  ``{fMRI} of visual encoding: reproducibility of activation,'' \emph{Human
  Brain Mapping}, vol.~9, pp. 156--64, 2000.

\bibitem{jaccard1901}
P.~Jaccard, ``\`{E}tude comparative de la distribution florale dans une portion
  des alpes et des jura,'' \emph{Bulletin del la Soci\`{e}t\`{e} Vaudoise des
  Sciences Naturelles}, vol.~37, p. 547–579, 1901.

\bibitem{biswaletal96}
\BIBentryALTinterwordspacing
B.~Biswal, A.~E. DeYoe, and J.~S. Hyde, ``Reduction of physiological
  fluctuations in {fMRI} using digital filters.'' \emph{Magn Reson Med},
  vol.~35, no.~1, pp. 107--113, January 1996. [Online]. Available:
  \url{http://view.ncbi.nlm.nih.gov/pubmed/8771028}
\BIBentrySTDinterwordspacing

\bibitem{genoveseetal97}
C.~R. Genovese, D.~C. Noll, and W.~F. Eddy, ``Estimating test-retest
  reliability in functional {MR} imaging: {I. S}tatistical methodology,''
  \emph{Magnetic Resonance in Medicine}, vol.~38, pp. 497--507, 1997.

\bibitem{hajnaletal94}
J.~V. Hajnal, R.~Myers, A.~Oatridge, J.~E. Schweiso, J.~R. Young, and G.~M.
  Bydder, ``Artifacts due to stimulus-correlated motion in functional imaging
  of the brain,'' \emph{Magnetic Resonance in Medicine}, vol.~31, pp. 283--291,
  1994.

\bibitem{maitraetal02}
R.~Maitra, S.~R. Roys, and R.~P. Gullapalli, ``Test-retest reliability
  estimation of functional mri data,'' \emph{Magnetic Resonance in Medicine},
  vol.~48, pp. 62--70, 2002.

\bibitem{chenandsmall07}
\BIBentryALTinterwordspacing
E.~E. Chen and S.~L. Small, ``Test-retest reliability in {fMRI} of language:
  Group and task effects,'' \emph{Brain and Language}, vol. 102, no.~2, pp.
  176--85, 2007. [Online]. Available:
  \url{http://dx.doi.org/10.1016/j.bandl.2006.04.015}
\BIBentrySTDinterwordspacing

\bibitem{buchsbaumetal05}
B.~R. Buchsbaum, S.~Greer, W.~L. Chang, and K.~F. Berman, ``Meta-analysis of
  neuroimaging studies of the wisconsin card-sorting task and component
  processes,'' \emph{Human Brain Mapping}, vol.~25, pp. 35--45, 2005.

\bibitem{derrfussetal05}
J.~Derrfuss, M.~Brass, J.~Neumann, and D.~Y. von Cramon, ``Involvement of
  inferior frontal junction in cognitive control: meta-analyses of switching
  and stroop studies,'' \emph{Human Brain Mapping}, vol.~25, pp. 22--34, 2005.

\bibitem{ridderinkhofetal04}
K.~R. Ridderinkhof, M.~Ullsperger, E.~A. Cron, and S.~Nieuwenhuis, ``The role
  of the medial frontal cortex in cognitive control,'' \emph{Science}, vol.
  306, pp. 443--447, 2004.

\bibitem{uttal01}
W.~R. Uttal, \emph{The new phrenology: the limits of localizing cognitive
  processes in the brain}.\hskip 1em plus 0.5em minus 0.4em\relax Cambridge,
  MA: The MIT Press, 2001.

\bibitem{woodetal98}
R.~P. Wood, S.~T. Grafton, J.~D.~G. Watson, N.~L. Sicotte, and J.~C. Mazziotta,
  ``Automated image registration. ii. intersubject validation of linear and
  non-linear models,'' \emph{Journal of Computed Assisted Tomography}, vol.~22,
  pp. 253--265, 1998.

\bibitem{mcgonigleetal00}
D.~J. McGonigle, A.~M. Howseman, B.~S. Athwal, K.~J. Friston, R.~S.~J.
  Frackowiak, and A.~P. Holmes, ``Variability in {fMRI}: an examination of
  intersession differences,'' \emph{Neuroimage}, vol.~11, pp. 708--734, 2000.

\bibitem{nolletal97}
F.~C. Noll, C.~R. Genovese, L.~E. Nystrom, A.~L. Vazquez, S.~D. Forman, W.~F.
  Eddy, and J.~D. Cohen, ``Estimating test-retest reliability in functional
  {MR} imaging. {II. A}pplication to motor and cognitive activation studies,''
  \emph{Magnetic Resonance in Medicine}, vol.~38, pp. 508--517, 1997.

\bibitem{weietal04}
X.~Wei, S.~S. Yoo, C.~C. Dickey, K.~H. Zou, C.~R. Guttman, and L.~P. Panych,
  ``Functional {MRI} of auditory verbal working memory: long-term
  reproducibility analysis,'' \emph{Neuroimage}, vol.~21, pp. 1000--1008, 2004.

\bibitem{shroutandfleiss79}
P.~E. Shrout and J.~L. Fleiss, ``Intraclass correlations: Uses in assessing
  rater reliability,'' \emph{Psychological Bulletin}, vol.~86, no.~2, pp.
  420--428, 1979.

\bibitem{koch82}
G.~G. Koch, ``Intraclass correlation coefficient,'' in \emph{Encyclopedia of
  Statistical Sciences}, S.~Kotz, , and N.~L. Johnson, Eds., vol.~4.\hskip 1em
  plus 0.5em minus 0.4em\relax New York: John Wiley and Sons, 1982, pp.
  213--217.

\bibitem{mcgrawandwong96}
K.~O. McGraw and S.~P. Wong, ``Forming inferences about some intraclass
  correlation coefficients,'' \emph{Psychological Methods}, vol.~1, no.~1, pp.
  30--46, 1996.

\bibitem{aronetal04}
A.~R. Aron, D.~Shohamy, J.~Clark, C.~Myers, M.~A. Gluck, and R.~A. Poldrack,
  ``Human midbrain sensitivity to cognitive feedback and uncertainty during
  classification learning,'' \emph{Journal of Neurophysiology}, vol.~92, pp.
  1144--1152, 2004.

\bibitem{fernandezetal03}
G.~Fern\'{a}ndez, K.~Sprecht, S.~Weis, I.~Tendolkar, M.~Reuber, J.~Fell, K.~P.,
  J.~Ruhlmann, J.~Reul, and C.~E. Elger, ``Intrasubject reproducibility of
  presurgical language lateralization and mapping using {fMRI},''
  \emph{Neurology}, vol.~60, pp. 969--975, 2003.

\bibitem{friedmanetal08}
\BIBentryALTinterwordspacing
L.~Friedman, H.~Stern, G.~G. Brown, D.~H. Mathalon, J.~Turner, G.~H. Glover,
  R.~L. Gollub, J.~Lauriello, K.~O. Lim, T.~Cannon, D.~N. Greve, H.~J.
  Bockholt, A.~Belger, B.~Mueller, M.~J. Doty, J.~He, W.~Wells, P.~Smyth,
  S.~Pieper, S.~Kim, M.~Kubicki, M.~Vangel, and S.~G. Potkin, ``Test-retest and
  between-site reliability in a multicenter {fMRI} study,'' \emph{Human Brain
  Mapping}, vol.~29, no.~8, pp. 958--972, 2008. [Online]. Available:
  \url{http://dx.doi.org/10.1002/hbm.20440}
\BIBentrySTDinterwordspacing

\bibitem{manoachetal01}
D.~S. Manoach, E.~F. Halpern, T.~S. Kramer, Y.~Chang, D.~C. Goff, S.~L. Rauch,
  D.~N. Kennedy, and R.~L. Gollub, ``Test-retest reliability of a functional
  {MRI} working memory paradigm in normal and schizophrenic subjects,''
  \emph{American Journal of Psychiatry}, vol. 158, pp. 955--958, 2001.

\bibitem{miezinetal00}
F.~M. Miezin, L.~Maccotta, J.~M. Ollinger, S.~E. Petersen, and R.~L. Buckner,
  ``Characterizing the hemodynamic response: effects of presentation rate,
  sampling procedure, and the possibility of ordering brain activity based on
  relative timing,'' \emph{Neuroimage}, vol.~11, pp. 735--759, 2000.

\bibitem{raemekersetal07}
M.~Raemekers, M.~Vink, B.~Zandbelt, R.~J.~A. van Wezel, R.~S. Kahn, and N.~F.
  Ramsey, ``Test-retest reliability of {fMRI} activation during prosaccades and
  antisaccades,'' \emph{Neuroimage}, vol.~36, pp. 532--542, 2007.

\bibitem{sprechtetal03}
K.~Sprecht, K.~Willmes, N.~J. Shah, and L.~J\"{a}ncke, ``Assessment of
  reliability in functional imaging studies,'' \emph{Journal of Magnetic
  Resonance Imaging}, vol.~17, pp. 463--471, 2003.

\bibitem{dice45}
L.~R. Dice, ``Measures of the amount of ecologic association between species,''
  \emph{Ecology}, vol.~26, pp. 297--302, 1945.

\bibitem{sorensen48}
T.~S{\o}rensen, ``A method of establishing groups of equal amplitude in plant
  sociology based on similarity of species and its application to analyses of
  the vegetation on danish commons,'' \emph{Biologiske Skrifter / Kongelige
  Danske Videnskabernes Selskab}, vol.~5, no.~4, p. 1–34, 1948.

\bibitem{tulloss97}
R.~E. Tulloss, ``Assessment of similarity indices for undesirable properties
  and a new tripartite similarity index based on cost functions,'' in
  \emph{Mycology in Sustainable Development: Expanding Concepts, Vanishing
  Borders}, M.~E. Palm and I.~H. Chapela, Eds.\hskip 1em plus 0.5em minus
  0.4em\relax North Carolina: Parkway Publishers, 1997, pp. 122--143.

\bibitem{ruddelletal07}
S.~Ruddell, S.~Twiss, and P.~Pomeroy, ``Measuring opportunity for sociality:
  quantifying social stability in a colonially breeding phocid,'' \emph{Animal
  Behaviour}, vol.~74, pp. 1357--1368, 2007.

\bibitem{levandowskyandwinter71}
M.~Levandowsky and D.~Winter, ``Distance between sets,'' \emph{Nature}, vol.
  234, pp. 34--35, 1971.

\bibitem{kherifetal03}
F.~Kherif, J.-B. Poline, S.~Meriaux, H.~Benali, G.~Flandin, and M.~Brett,
  ``Group analysis in functional neuroimaging: selecting subjects using
  similarity measures,'' \emph{Neuroimage}, vol.~20, no.~4, pp. 2197--2208,
  2003.

\bibitem{luoandnichols03}
W.-L. Luo and T.~E. Nichols, ``Diagnosis and exploration of massively
  univariate neuroimaging models,'' \emph{Neuroimage}, vol.~19, no.~3, pp.
  1014--1032, 2003.

\bibitem{seghieretal07}
M.~Seghier, K.~Friston, and C.~Price, ``Detecting subject-specific activations
  using fuzzy clustering,'' \emph{Neuroimage}, vol.~36, pp. 594--605, 2007.

\bibitem{mcnameeandlazar04}
R.~L. McNamee and N.~A. Lazar, ``Assessing the sensitivity of {fMRI} group
  maps,'' \emph{Neuroimage}, vol.~22, no.~2, pp. 920--931, 2004.

\bibitem{woolrich08}
M.~Woolrich, ``Robust group analysis using outlier inference,''
  \emph{Neuroimage}, vol.~41, pp. 286--301, 2008.

\bibitem{colwellandcoddington94}
R.~K. Colwell and J.~A. Coddington, ``Estimating terrestrial biodiversity
  through extrapolation,'' \emph{Philosophical Transactions of the Royal
  Society of London}, vol. 345, pp. 101--118, 1994.

\bibitem{bapatandraghavan97}
R.~B. Bapat and T.~E.~S. Raghavan, \emph{Nonnegative Matrices and
  Applications}.\hskip 1em plus 0.5em minus 0.4em\relax Cambridge, United
  Kingdom: Cambridge University Press, 1997.

\bibitem{efron79}
B.~Efron, ``Bootstrap methods: another look at the jackknife,'' \emph{Annals of
  Statistics}, vol.~7, pp. 1--26, 1979.

\bibitem{efronandgong83}
B.~Efron and G.~Gong, ``A leisurely look at the bootstrap, the jackknife, and
  cross-validation,'' \emph{American Statistician}, vol.~37, no.~1, pp. 36--48,
  1983.

\bibitem{benjaminiandhochberg95}
\BIBentryALTinterwordspacing
Y.~Benjamini and Y.~Hochberg, ``Controlling the false discovery rate: A
  practical and powerful approach to multiple testing,'' \emph{Journal of the
  Royal Statistical Society. Series B (Methodological)}, vol.~57, no.~1, pp.
  289--300, 1995. [Online]. Available: \url{http://dx.doi.org/10.2307/2346101}
\BIBentrySTDinterwordspacing

\bibitem{allisonetal95}
D.~B. Allison, F.~Paultre, M.~I. Goran, E.~T. Poehlman, and S.~B. Heymseld,
  ``Statistical considerations regarding the use of ratios to adjust data,''
  \emph{International Journal of Obesity}, vol.~19, pp. 644--652, 1995.

\bibitem{coxandhyde97}
R.~W. Cox and J.~S. Hyde, ``Software tools for analysis and visualization of
  {fMRI} data,'' \emph{NMR in Biomedicine}, vol.~10, no. 4-5, pp. 171--178,
  1997.

\bibitem{adler81}
R.~Adler, \emph{The geometry of random fields}.\hskip 1em plus 0.5em minus
  0.4em\relax New York: Wiley, 1981.

\bibitem{worsley94}
K.~J. Worsley, ``Local maxima and the expected euler characteristic of
  excursion sets of $\chi^2$, $f$ and $t$ fields,'' \emph{Advances in Applied
  Probability}, vol.~26, pp. 13--42, 1994.

\bibitem{R}
\BIBentryALTinterwordspacing
{R Development Core Team}, \emph{R: A Language and Environment for Statistical
  Computing}, R Foundation for Statistical Computing, Vienna, Austria, 2009,
  {ISBN} 3-900051-07-0. [Online]. Available: \url{http://www.R-project.org}
\BIBentrySTDinterwordspacing

\bibitem{lazaretal02}
N.~A. Lazar, B.~Luna, J.~A. Sweeney, and W.~F. Eddy, ``Combining brains: a
  survey of methods for statistical pooling of information,''
  \emph{Neuroimage}, vol.~15, pp. 538--50, 2002.

\end{thebibliography}
\end{document}